\documentclass[11pt]{article}
\pdfoutput=1
\usepackage[T1]{fontenc}
\usepackage[utf8]{inputenc}
\usepackage{authblk}

\usepackage{graphicx}
\usepackage{slashed}
\usepackage{feynmp}
\usepackage{multirow}
\usepackage{epstopdf}
\usepackage[body={17.5cm, 21cm},right=2cm]{geometry}
\usepackage{amssymb}
\usepackage{amsmath}
\usepackage{caption}
\usepackage{subcaption}
\usepackage{verbatim}
\usepackage{authblk}
\usepackage{booktabs}
\usepackage{fancyhdr}
\usepackage{ctable}
\usepackage{feynmp}
\usepackage{color}
\numberwithin{equation}{section}

\newcommand\eea{\end{eqnarray}}
\newcommand\bea{\begin{eqnarray}}

\def\beq{\begin{equation}}
\def\eeq{\end{equation}}

\newcommand{\be}{\begin{equation}}
\newcommand{\ee}{\end{equation}}
\newcommand{\ba}{\begin{align}}
\newcommand{\ea}{\end{align}}
\newcommand{\bg}{\begin{gather}}
\newcommand{\eg}{\end{gather}}
\newcommand{\bseq}{\begin{subequations}}
\newcommand{\eseq}{\end{subequations}}

\newcommand{\GeV}{\,\mathrm{GeV}}

\newcommand{\fb}{\,\mathrm{fb}}
\newcommand{\ifb}{\mathrm{fb}^{-1}}
\newcommand{\pb}{\,\mathrm{pb}}

\begin{document}

\begin{titlepage}
%
%
%
%
\title{Minimal Signatures of Naturalness}
\author[1]{Sonia El Hedri}
\author[2]{Anson Hook}
\affil[1]{SLAC, Stanford University, Menlo Park, CA 94025 USA}
\affil[2]{Institute for Advanced Studies, Princeton University, Princeton, NJ 08544 USA}
\maketitle

\begin{abstract}

We study the naturalness problem using a model independent bottom up approach by considering models where only the interaction terms needed to cancel the Higgs quadratic divergences are present.
If quadratic divergences are canceled by terms linear in the Higgs field, then the collider phenomenology is well covered by current electroweakino and fourth generation searches.  
If quadratic divergences are canceled by terms bilinear in the Higgs field, then the signatures are highly dependent on the quantum numbers of the new particles.  Precision Higgs measurements can reveal the presence of new particles with either vevs or Standard Model charges.
If the new particles are scalar dark matter candidates, their direct and indirect detection signatures will be highly correlated and within the reach of XENON100 and Fermi.  Observation at one of these experiments would imply observation at the other one.
Observable LHC decay channels can also arise if the new particles mix with lighter states.
This decay channel involves only the Higgs boson and not the gauge bosons.  Observation of such decays would give evidence that the new particle is tied to the naturalness problem.

\end{abstract}

\vspace{1cm}

\end{titlepage}

\section{Introduction}

\begin{figure}[b]
    \begin{align*}
        \parbox{32mm}{
        \begin{fmffile}{gluon4}
        \begin{fmfgraph}(80,60)
            \fmfkeep{fermion}
            \fmfleft{i} 
            \fmfright{o} 
            \fmf{dashes}{i,v1} 
            \fmf{dashes}{v2,o}
            \fmf{fermion,left,tension=.3}{v1,v2,v1}
            \end{fmfgraph}
    \end{fmffile}
} \propto - m_t^2 \Lambda^2\quad\quad\quad\quad\quad\quad
\parbox{32mm}{
        \begin{fmffile}{gauge}
        \begin{fmfgraph}(80,60)
            \fmfkeep{fermion}
            \fmfleft{i} 
            \fmfright{o} 
            \fmf{dashes}{i,v1} 
            \fmf{dashes}{v1,o}
            \fmf{photon,right,tension=.6}{v1,v1}
            \end{fmfgraph}
    \end{fmffile}
}\propto m_V^2\Lambda^2
\end{align*}
\caption{Top quark (left) and gauge boson (right) contributions to the one-loop quadratic divergences of the Higgs mass. $\Lambda$ is the cutoff scale.\label{Fig: div}}
\end{figure}
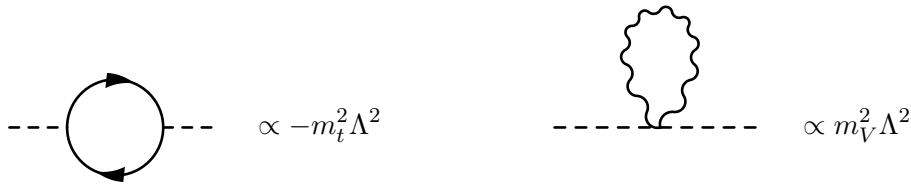

The LHC has recently discovered a Higgs like boson with mass $125\GeV$ \cite{Chatrchyan:2012tx,ATLAS:2012ae}. In the Standard Model (SM),  the Higgs boson receives quadratically divergent contributions to its mass from radiative corrections. The biggest contributions to these quadratic divergences come from top quark and gauge boson loops and are shown in Fig~\ref{Fig: div}. Such radiative corrections will depend on the new physics scale, being proportional to the square of the cutoff scale. If new physics appears only at the GUT scale, these radiative corrections to the bare Higgs mass will be of the order of $10^{32}\GeV$. The observed Higgs mass would then result from a miracle cancellation between the bare Higgs mass and the radiative corrections, requiring a large amount of fine tuning of UV parameters. Various models such as supersymmetry, extra dimensions \cite{ArkaniHamed:1998rs,Randall:1999ee}, Little Higgs \cite{ArkaniHamed:2001nc,ArkaniHamed:2002qy}, Twin Higgs \cite{Chacko:2005pe} and folded SUSY \cite{Burdman:2006tz} have all been proposed to solve this naturalness issue, but the existence of other possible solutions should not be ignored.  The current solutions give rise to very diverse signatures at colliders and dark matter detectors.  However, most of these signatures are uncorrelated with the cancelation of the Higgs quadratic divergence.

  There have been bottom up approaches to naturalness in the context of SUSY \cite{Carmi:2012yp,Hall:2011aa,Brust:2011tb,Papucci:2011wy,Kats:2011qh} or Little Higgs \cite{Perelstein:2003wd,Berger:2012ec,Fajfer:2013wca}.  However, as mentioned before, the phenomenology of these models is not necessarily tied to the cancelation of the Higgs quadratic divergences.   Applying Occam's razor to models of naturalness, the simplest solution of the naturalness problem would cancel quadratic divergences without any additional interactions or matter content\footnote{Anthropics may favor using additional quartic interactions to fine tune the Higgs mass rather than using the bare mass.
As generic particles have no other interactions with the SM, the resulting theory cancels quadratic divergences with no additional interactions.}.

    Thus in this paper we focus on the signatures of interactions necessary for the cancelation of quadratic divergences. These signatures vanish in the limit where the quadratic divergences are not canceled.  This approach leads us to consider the only two terms in the Lagrangian which contribute to the 1-loop quadratic divergences of the Higgs mass,  
\bea
\label{Eq: rare decay}
\mathcal{L} \supset \lambda_1 H \psi_i \psi_j + \lambda_2 \psi_i^\dagger \psi_i HH^\dagger
\eea
where the quantum numbers of the fields $\psi_i$ are only constrained by gauge invariance.

Ideally, these couplings would be measured at a collider and shown to cancel the quadratic divergences induced by the SM fields.  $\lambda_1$ can be directly observed through the production and decay of  $\psi_i$.  However, even proving the existence of $\lambda_2$ is difficult, since it is not directly related to the production or decay of the $\psi_i$.  Direct observation at a low luminosity LHC is not feasible.  Depending on the quantum numbers of $\psi_i$, this term may be the only renormalizable interaction, preventing any other means of detection.  

This article studies the signatures of models where either $\lambda_1$ or $\lambda_2$ are used to cancel quadratic divergences.  Models which use $\lambda_1$ to cancel quadratic divergences have signatures that are similar to MSSM electroweak phenomenology \cite{Alves:2011wf} or fourth generation models \cite{CMS:2012pva,Nektarijevic:2012zf}.  Models which use $\lambda_2$ to cancel quadratic divergences can have interesting correlated dark matter direct and indirect signatures.  The $\lambda_2$ term can also lead to up to $10\%$ modifications of the Higgs production cross section or of its decay width to gauge bosons.

Predicting other, more visible, collider signatures coming from the $\lambda_2$ term requires introducing additional terms in the Lagrangian.  Generically, these terms would lead to signatures which are not directly related to the quartic term of Eq.~\ref{Eq: rare decay}.  Introducing mass mixing with a lighter state is the only way of getting an observable decay channel for $\psi_i$ without making additional assumptions on the UV symmetry enforcing naturalness. If $\psi_i$ mixes with a lighter field, then $\lambda_2$ induces a decay of $\psi_i$ to this field and a Higgs boson.


This article is organized as follows.  Sec.~\ref{Sec: linear} analyzes the signatures of simplified models which use a coupling linear in the Higgs field to cancel quadratic divergences.  Sec.~\ref{Sec: bilinear} studies simplified models which use a coupling bilinear in the Higgs field to cancel quadratic divergences.
Sec.~\ref{Sec: bilinear indirect} studies signatures that result from mass mixing and the bilinear term in Eq.~\ref{Eq: rare decay}.  We conclude in Sec.~\ref{Sec: conclusion}.

\section{Simplified models with couplings linear in H}
\label{Sec: linear}

We consider minimal extensions of the SM where there is only the term 
\bea
\label{Eq: linear}
\mathcal{L} \supset \lambda H \bar{\psi}_1 \psi_2
\eea
in the Lagrangian.  In order for this coupling to contribute to the Higgs quadratic divergence at 1-loop, we need $\psi_1$ and $\psi_2$ to be fermions.  For a hard UV cutoff, the 1-loop quadratically divergent contribution to the Higgs mass is always negative and is $-\frac{|\lambda|^2 \Lambda^2}{8 \pi^2}$.  Thus if this term in the Lagrangian is to cancel quadratic divergences, it must cancel the positive divergences due to the gauge bosons. Due to gauge invariance, either $\psi_1$ or $\psi_2$ has to be charged under $SU(2)$.  The minimal models considered in this section are labeled by whether $\psi_2$ is a SM particle or not.  Other non-minimal signatures can be obtained by combining these models.
\begin{table}
    \centering
    \begin{tabular}{ccc}\toprule
        $\psi_{1,2}$ stable new particle & Production Channel & Signatures \\
        \midrule
        Yes  & pair production & 2 V/H + R-hadrons/CHAMPS/$\slashed{E}_T$ \\
        No & pair + single production  & 2 V/H + leptons/tops/jets\\
 & & V/H + leptons/tops/jets\\
        \bottomrule
    \end{tabular}
    \caption{Possible signatures associated with minimal models which use the interaction term in Eq.~\ref{Eq: linear} to cancel Higgs quadratic divergences.  In both cases particles are produced through gluons/gauge bosons.}
\end{table}

\paragraph{If both $\psi_1$ and $\psi_2$ are non-SM particles,}they are pair produced through intermediate gauge bosons and then decay through gauge bosons and Higgses. In general, the associated decay pattern closely resembles the SUSY signatures involving the gauginos \cite{Alves:2011wf}. If the lighter of the $\psi$ is colored or charged, then we can have the pair production of R-hadrons \cite{Aad:2011yf,Aad:2012pra,Farrar:1978xj}, CHAMPS \cite{Chatrchyan:2013oca,Aad:2011mb} or electroweakino-like decays ending in R-hadrons or CHAMPS.  The LHC signatures of this scenario are decays that involve gauge and/or Higgs bosons and either missing $E_T$, CHAMPS or R-hadrons.  The similarity of these signatures with electroweakino decays is not an accident as these are the couplings that SUSY theories use to cancel the gauge quadratic divergences.

\paragraph{If only $\psi_1$ is a new particle,}
it can be both pair produced and singly produced.  Its pair production cross section is determined by its gauge quantum numbers while its single production cross section is determined by what standard model particle $\psi_1$ decays into. e.g. if $\psi_1$ is charged electroweakly and decays into leptons, then it is both pair and singly produced through intermediate W/Zs.  Since canceling the gauge quadratic divergences requires gauge and Yukawa couplings to be of the same order, both single and pair production matrix elements are comparable.  Because $\psi_1$ is heavier than its Standard Model partner, we will see single production before pair production.  For $\psi_1$ which are colored, the pair production through gluons may be seen before the single production through W/Zs.  The cross section for single production depends on $\lambda$ in Eq.~\ref{Eq: linear} so that its magnitude can be used to determine which gauge divergence is canceled.  The decay of $\psi_1$ is the same as in other fourth generation models \cite{CMS:2012pva,Nektarijevic:2012zf} or models with new vector-like fermions such as Little Higgs models  \cite{ArkaniHamed:2001nc,ArkaniHamed:2002qy}.

If $\psi_1$ is lighter than $\psi_2$, then the SM particle $\psi_2$ can have new three body decay modes. The cases of $\psi_1$ being colored or having an electroweak charge are both ruled out by the R-hadron \cite{Aad:2011yf,Aad:2012pra,Farrar:1978xj} and LEP bounds \cite{Achard:2001qw} respectively. If $\psi_1$ is a SM singlet, then it is neutrino like and after the Higgs obtains a vev,  Eq.~\ref{Eq: linear} generates a large mass term for the neutrinos.  Thus we find that without making additional assumptions, this scenario is also excluded.

\section{Simplified models with couplings bilinear in H}  
\label{Sec: bilinear}

In this section, we consider couplings of the form
\bea
\label{Eq: bilinear}
\mathcal{L} &\supset& \lambda \psi^\dagger \psi HH^\dagger \\
\label{Eq: bilinear2}
  &=&  \lambda \psi^\dagger \psi \frac{h^2}{2} +  \lambda v \psi^\dagger \psi h + \lambda \frac{v^2}{2} \psi^\dagger \psi 
\eea
 where the Higgs vev $v$ will be defined as
\begin{align*}
    v = 246\GeV
\end{align*}
all throughout this article.

For a hard UV cutoff, the quadratic divergence can be either positive or negative depending on the sign of $\lambda$.  Thus we have the option of canceling any quadratic divergence induced by either gauge bosons or fermions.

$\psi$ is either a scalar or a vector-like fermion but does not have to be charged under the SM. If $\psi$ is lighter than half the mass of the Higgs, the second term of Eq.~\ref{Eq: bilinear2} opens a new decay channel for the Higgs.  If $\psi$ is a scalar, it can get a vev and mix with the Higgs after EWSB.  If $\psi$ is charged under the SM, then it modifies the Higgs couplings to the gauge bosons at 1-loop.  Finally, if $\psi$ is a SM singlet, then it is a dark matter candidate and can be accessible at the current dark matter detection experiments.   All these possibilities give rise to various signatures, summarized in Tab.~\ref{Tab: bilinear}.


\subsection{$m_\psi < \frac{m_h}{2}$}
If the new particles are lighter than half the Higgs mass, then they can give rise to Higgs alternate decay modes. Light stable standard model singlets or neutral components of standard model triplets will lead to invisible Higgs decay modes. If these particles cancel top or gauge quadratic divergences then the decay widths into these new particles would be orders of magnitude above the decay width into bottoms.  Current LHC searches rule out decays widths this large \cite{ATLAS:2013pma}.  Canceling other quadratic divergences yields modifications too small to observe.

 Light electrically charged of colored particles have been ruled out by LEP \cite{Achard:2001qw}  and R-hadron searches respectively. Decays of new light $SU(2)$ charged particles to standard model states would lead to model dependent modifications of the Higgs branching ratios with respect to their standard model values. If the final states associated to $h\rightarrow\psi\psi$ do not significantly contribute to any of the main Higgs search channels, all the detection rates in these channels will be uniformely suppressed. In this last case, current LHC Higgs coupling measurements rule out the possibility of cancelling the top or gauge quadratic divergences \cite{ATLAS:2013sla,CMS:yva}.

\subsection{$\langle \psi \rangle \ne 0$}
\label{Sec: vevs}

This subsection considers the case where the new particle which cancels quadratic divergences is a scalar that obtains a vev.  If this scalar is an $SU(2)$ doublet, then the model is a two Higgs doublet model that satisfies the Veltman conditions \cite{Veltman:1980mj}. If the new particle is an SM singlet, the minimal Lagrangian is 
\bea
\label{Eq: hdm1}
\mathcal{L} &=& \frac{\lambda_1}{2} \left(H^\dagger H-\frac{v^2}{2}\right)^2 + \frac{\lambda_2}{2} \left(\Phi^\dagger\Phi - \frac{v_\Phi^2}{2}\right)^2 + \lambda \left(H H^\dagger - \frac{v^2}{2}\right) \left(\Phi \Phi^\dagger - \frac{v_\Phi^2}{2}\right)\\
&=& \lambda v v_\Phi h\phi + \frac{\lambda}{2} v_\Phi \phi h h + \frac{\lambda}{2} v \phi \phi h  + \frac{\lambda_1}{2}v hhh + \frac{\lambda_2}{2}v_\Phi \phi\phi\phi + \text{ mass terms } + \text{ quartic terms}
\label{Eq: hdm}
\eea

After both $H$ and $\Phi$ obtain a vev, Eq.~\ref{Eq: hdm} induces mass mixing.  In the simplest case where $\Phi$ interacts only through this term, its decays are determined by the mixing with the Higgs and by the three point term arising from Eq.~\ref{Eq: hdm}. The mass eigenstates are 
\bea
 \begin{pmatrix}
h_m \\  \phi_m
\end{pmatrix} =  \begin{pmatrix} \cos\alpha & \sin\alpha \\ -\sin\alpha & \cos\alpha \end{pmatrix}  \begin{pmatrix}
h \\  \phi
\end{pmatrix} 
\eea
This mixing suppresses all of the SM Higgs couplings by an amount $\cos\alpha$.  The mixing angle is constrained by precision measurements of the Higgs couplings. Because the current best fit values of these couplings in ATLAS are high compared to the SM values \cite{ATLAS:2013sla}, the corresponding 2$\sigma$ bound is tight and is
\bea
\cos\alpha \ge 0.93
\eea

The production of $\phi_m$ occurs only through mixing with the Higgs and is proportional to $\sin^2\alpha \le 14\%$.  If $\phi_m$ is lighter than twice the Higgs mass, it acts like a heavy SM Higgs boson with uniformly suppressed couplings; the signal strength is suppressed by $\sin^2\alpha$ in all channels.  If $\phi_m$ is heavier than twice the Higgs mass, $\phi_m$ can decay into 2 Higgses.  By the goldstone boson equivalence theorem, the branching ratios into Ws/Zs/hs is 2:1:1 in the large mass limit.  The best limit on heavy SM Higgs arises from $H \rightarrow ZZ \rightarrow llll$ \cite{ATLAS:2012bmv,CMS:yxa}. The bounds on $\sin^2\alpha$ are generally of about $20$\% and can be as low as $10$\% for very specific masses.  Thus we find that precision Higgs physics is the best method to constrain this class of models.

\begin{table}[t]
\centering
\begin{tabular}{cc}\toprule
Scenario&Signature
\\
\midrule
$m_\psi < \frac{m_h}{2}$ & Higgs decays to missing $E_T$, CHAMPS, R-hadrons \\
$\langle \psi \rangle \ne 0$  & suppressed Higgs couplings and a heavy Higgs-like scalar \\
$\psi$ charged under SM & $\mathcal{O}(10\%)$ changes to $H \rightarrow \gamma \gamma / g g$ \\
$\psi$ is dark matter & Correlated direct and indirect detection signatures \\
\bottomrule
\end{tabular}
\caption{
The scenarios and collider signatures that results from minimal models of naturalness which cancel the quadratic divergences through the term shown in Eq.~\ref{Eq: bilinear}.  The quantum numbers and spin of $\psi$ are allowed to vary when not specified.
 }
\label{Tab: bilinear}
\end{table}

\subsection{$\psi$ is charged under the SM}
  If $\psi$ has SM quantum numbers, the three point couplings in Eq.~\ref{Eq: bilinear2} can lead the new particles running in loops to contribute to the Higgs decays to gauge bosons.  Measuring the deviation of these decay rates from their SM values gives information as to which quadratic divergence could be cancelled.  As an example, consider a singlet fermion which cancels the top quadratic divergence and has charge 1.  The relevant terms in the Lagrangian are
\bea
\mathcal{L} \supset -m \psi^\dagger \psi + \frac{3 y_t^2}{2 m} \psi^\dagger \psi HH^\dagger
\eea
The modification of the Higgs decay rate to two photons is given by the low energy theorem \cite{Ellis:1975ap,Shifman:1979eb} to be
\bea
\frac{\Gamma(h\rightarrow \gamma \gamma)}{\Gamma(h\rightarrow \gamma \gamma)_{\text{SM}}} = \left |  1 - \frac{1}{6.49} Q^2 \frac{4}{3} \left ( \frac{\partial\log m_\psi}{\partial\log v} \right ) \left (1 + \frac{7 m_h^2}{120 m_\psi^2} \right ) \right |^2
\eea
If this uncolored top partner had a mass of 600 $\GeV$, the diphoton decay width of the Higgs will be larger than in the standard model by 10.5\%. The current ATLAS and CMS $95\%$ confidence limits on the Higgs to diphoton rate allow excesses of up to $90\%$ for ATLAS and $20\%$ for CMS \cite{ATLAS:2013oma,CMS:photons}. Although $10\%$ modifications of the Higgs to diphoton rate should be within reach of the future LHC precision Higgs measurements, determining the Higgs couplings to other gauge bosons with a similar precision would take much longer.

\subsection{$\psi$ is a dark matter candidate}
\label{Sec: DM constraints}
If $\psi$ is a dark matter candidate, Eq.~\ref{Eq: bilinear2} shows that after EWSB, Higgs exchange with nuclei and annihilation into $H \rightarrow b \bar{b} \rightarrow \pi^0s$ giving many photons gives rise to direct and indirect detection signatures respectively.  The dark matter particle's coupling to the Higgs boson is
\bea
\mathcal{L} &=& \lambda H H^\dagger X X^\dagger \\
&=& \lambda \frac{h^2}{2} X X^\dagger + \lambda h v X X^\dagger
\eea
For fermions/complex scalars, we have 
\bea
\lambda_f = \frac{N_c y_t^2}{2N_f m_X} \qquad \lambda_s = 2 \frac{N_c}{N_s}  y_t^2
\eea
where $N_c$ is the color factor ($N_c=3$) and $N_s$, $N_f$ are the multiplicity of scalar and fermionic dark matter candidates.  Dark matter has been considered to interact using this term in other contexts \cite{Andreas:2008xy,Poland:2008ev,Bazzocchi:2012pp}.  For simplicity, we will be assuming a single complex dark matter field.

  \begin{figure}
\centering
\begin{tabular}{cc}
\begin{subfigure}{.5\textwidth}
  \centering
  \includegraphics[width=3.12in]{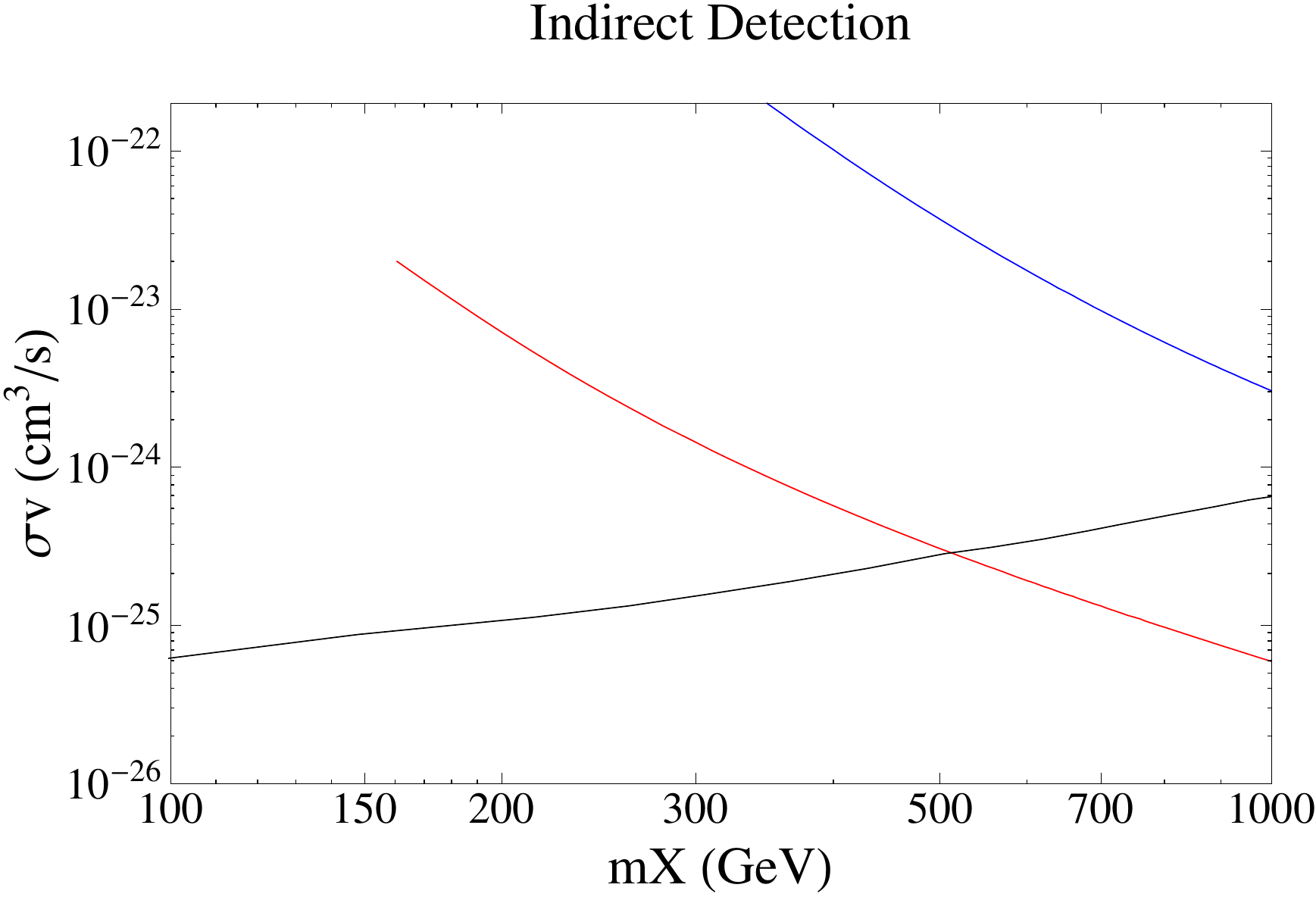}
  \label{fig:sub1}
\end{subfigure}
   &   \begin{subfigure}{.5\textwidth}
  \centering
\includegraphics[width=3.12in]{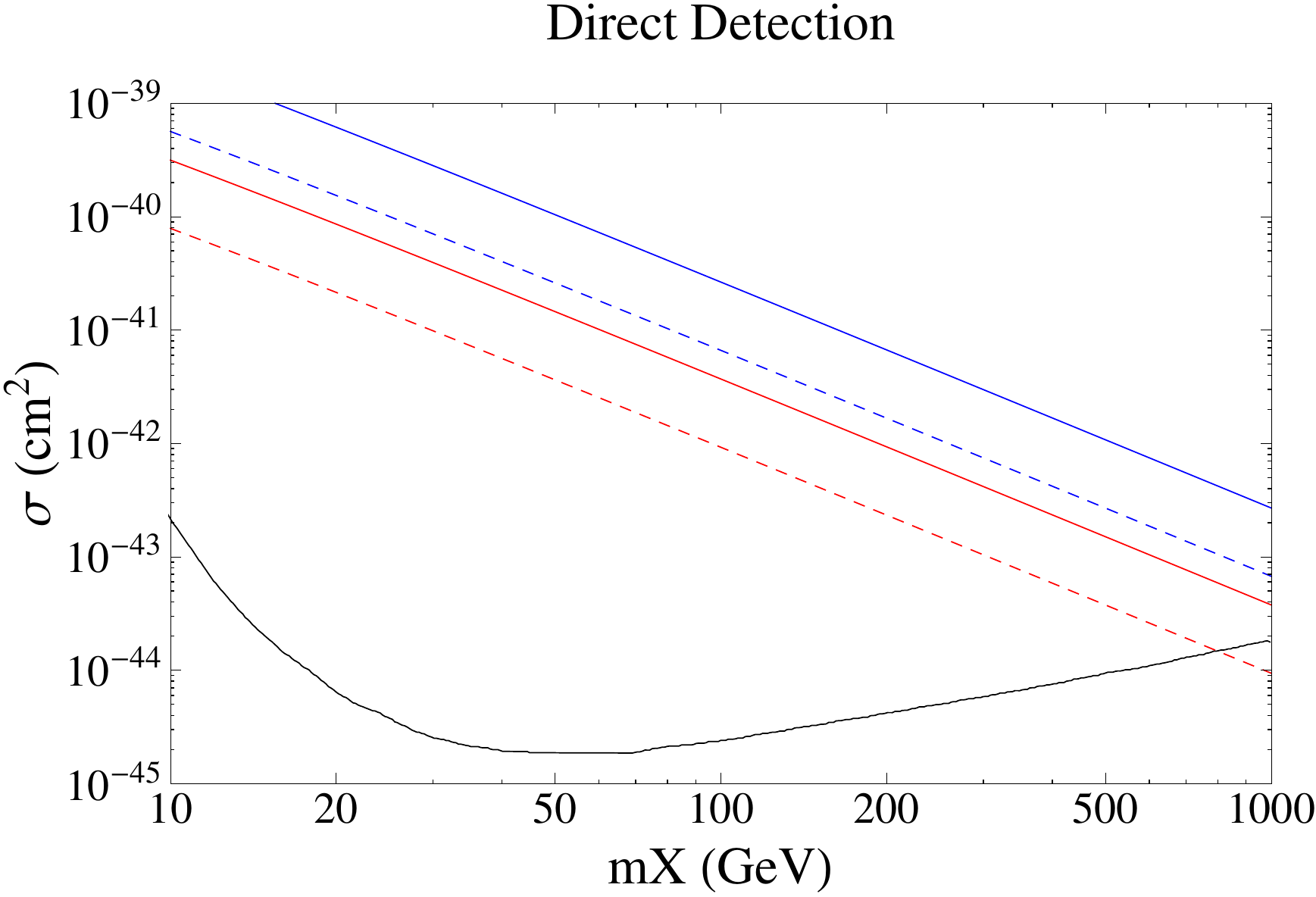}
  \label{fig:sub2}
\end{subfigure} \\
\multicolumn{2}{c}{
\begin{subfigure}{.5\textwidth}
  \centering
  \includegraphics[width=3.12in]{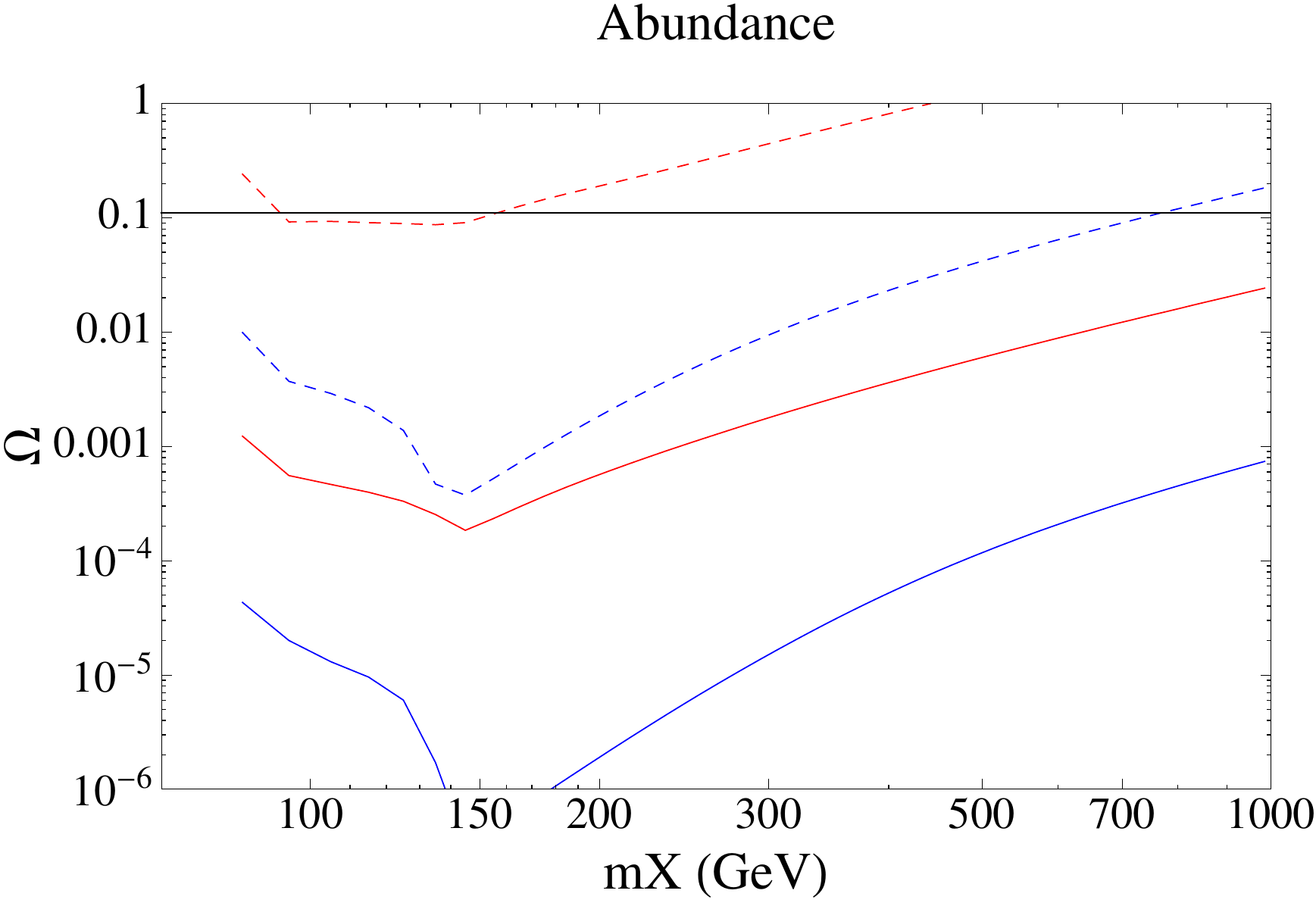}
  \label{fig:sub1}
\end{subfigure}}
\end{tabular}
\caption{The abundance, direct and indirect detection bounds on dark matter particles which cancels quadratic divergences.  The indirect detection bound is the Fermi bound on $X X \rightarrow b \bar{b}$ \cite{DrlicaWagner:2012ry}, while the annihilation channel is into two Higgses.  Using pythia, it was seen that the number and distributions of photons from a pair of Higgses and a pair of bottoms was within 20\% of each other.  Thus the plot can be used to find the approximate indirect detection bound.  The direct detection bound is taken from XENON100 \cite{Aprile:2012nq} and the relic abundance is taken from \cite{Eidelman:2004wy}.  Bounds are placed assuming $X$ makes up all of the dark matter.  The blue line signals a cancelation of the top quadratic divergence while a red line is the gauge quadratic divergence.  The black lines are the current bounds.  Solid lines are complex scalar dark matter while dashed lines are dirac fermion dark matter.}
\label{Fig: dark matter}
\end{figure}

Direct detection constrains the spin independent cross section
\bea
\sigma_{p,n, SI} &=& \frac{a}{\pi} \frac{m_p^2}{(m_X + m_p)^2} \frac{9 y_t^4 m_p^2}{m_h^4} f^2 \\
f &=& \frac{6}{27} + \frac{21}{27}(f_{Tu}+f_{Td}+f_{Ts})
\eea
where $a=4$ if $X$ is a real scalar, $a=1$ if $X$ is a majorana fermion or a complex scalar and $a=1/4$ if $X$ is a dirac fermion \cite{Fitzpatrick:2010em}. We used the values $f_{Tu}=f_{Td}=0.025$ and $f_{Ts}=0.0532$ as suggested by lattice studies \cite{Giedt:2009mr}. The results are shown in Fig.~\ref{Fig: dark matter} and show that sub-TeV singlet dark matter particles canceling the top quadratic divergences are excluded by XENON100. 

Continuum photons constrain the annihilation rate of dark matter into bottoms/Ws/Zs.  Fermi provided the bound on dark matter annihilation into a pair of bottoms \cite{DrlicaWagner:2012ry} while, in this particular model, dark matter annihilates into a pair of Higgses. Higgses dominantly decay into bottoms and Ws which in turn give many pions and hence many photons.  Using pythia, it was found that the number and distributions of the photons from dark matter annihilating into bottoms and Higgses were within 20\% of each other.  Therefore, we compare the dark matter annihilation rate into a pair of Higgses to the bound on dark matter annihilation into a pair of bottoms.  Fig.~\ref{Fig: dark matter} gives a rough estimate on what the exclusion limits should be.  microOMEGAs was used to calculate the annihilation cross sections.  For fermions, the vanishing of the cross section for annihilation into Higgs bosons can be understood because the $J=0$ initial state of two fermions has $CP = -1$ while the final state of two identical scalar particles cannot have spin 0 and $CP=-1$.  Thus the annihilation amplitude must vanish in the $v=0$ limit.


If the dark matter sector solves the Higgs naturalness problem, it leads to unique predictions.  Fermionic dark matter is visible in direct detection experiments but not in indirect detection experiments.  Scalar dark matter has a much more distinct signature, which is that the ratio of direct to indirect detection cross sections is
\bea
\frac{\sigma_{p,n, SI}}{\sigma v} = \frac{16 f^2 m_p^4}{m_h^4} = 1.5 \times 10^{-19} \frac{cm^2}{cm^3/s}
\label{Eq: correlation}
\eea
in the large dark matter mass limit.

 If scalar dark matter cancels the top quadratic divergences, it annihilates too efficiently so obtaining the correct relic abundance requires non-thermal cosmology.  Fermionic dark matter annihilates less efficiently as its cross sections are velocity suppressed so it can potentially give the correct thermal relic abundance.  Unfortunately, the parameter points which give the correct abundance are ruled out by direct detection constraints. Relic abundances were calculated using micrOMEGAs \cite{Belanger:2010pz}. 

As can be seen from Fig.~\ref{Fig: dark matter}, sub-TeV singlet scalar dark matter cannot cancel the top quadratic divergences; if it cancels gauge quadratic divergences, then its detection could be just around the corner for both direct and indirect detection experiments. Simultaneous detection in both experiments with cross sections obeying Eq.~\ref{Eq: correlation} would be strong evidence that dark matter is a scalar involved in the cancelation of quadratic divergences and has a non-thermal production mechanism.

 Direct evidence of new particles canceling the Higgs quadratic divergences can be obtained through dark matter direct and indirect detection experiments or through LHC precision Higgs measurements. Probing naturalness through precision Higgs measurements will take many years. Some of the fields $\psi$ in Eq.~\ref{Eq: bilinear} have also none of the characteristics mentioned in Tab.~\ref{Tab: bilinear}. Such fields would have no visible signatures in current experiments if only Eq.~\ref{Eq: bilinear} is present.  While as shown in \cite{Craig:2013xia}, future experiments such as the ILC can probe naturalness, the LHC has limited reach. In order to obtain visible signatures at the LHC, one is led to tying the cancelation of quadratic divergences to the decays of new particles.

\section{LHC signatures with minimal assumptions}
\label{Sec: bilinear indirect}
  
  In any theory where the Higgs quadratic divergences cancel, the bilinear term 
  \bea
  \label{Eq: bilinear yet again}
\mathcal{L} &\supset& \lambda \psi^\dagger \psi HH^\dagger
\eea
is related to the Yukawa couplings by a symmetry.  If a specific model is assumed, then the structure associated to this symmetry can be tested to obtain indirect evidence for the cancelation of quadratic divergences.  e.g. in SUSY the gauginos have gauge strength trilinear interactions or Higgsinos having SM Yukawa strength trilinear interactions.  Instead, motivated by our bottom up approach, we are lead to look for IR effects independent of the UV symmetry which can shed light on the bilinear term.

  \begin{figure}
\centering
  \includegraphics[width=3in]{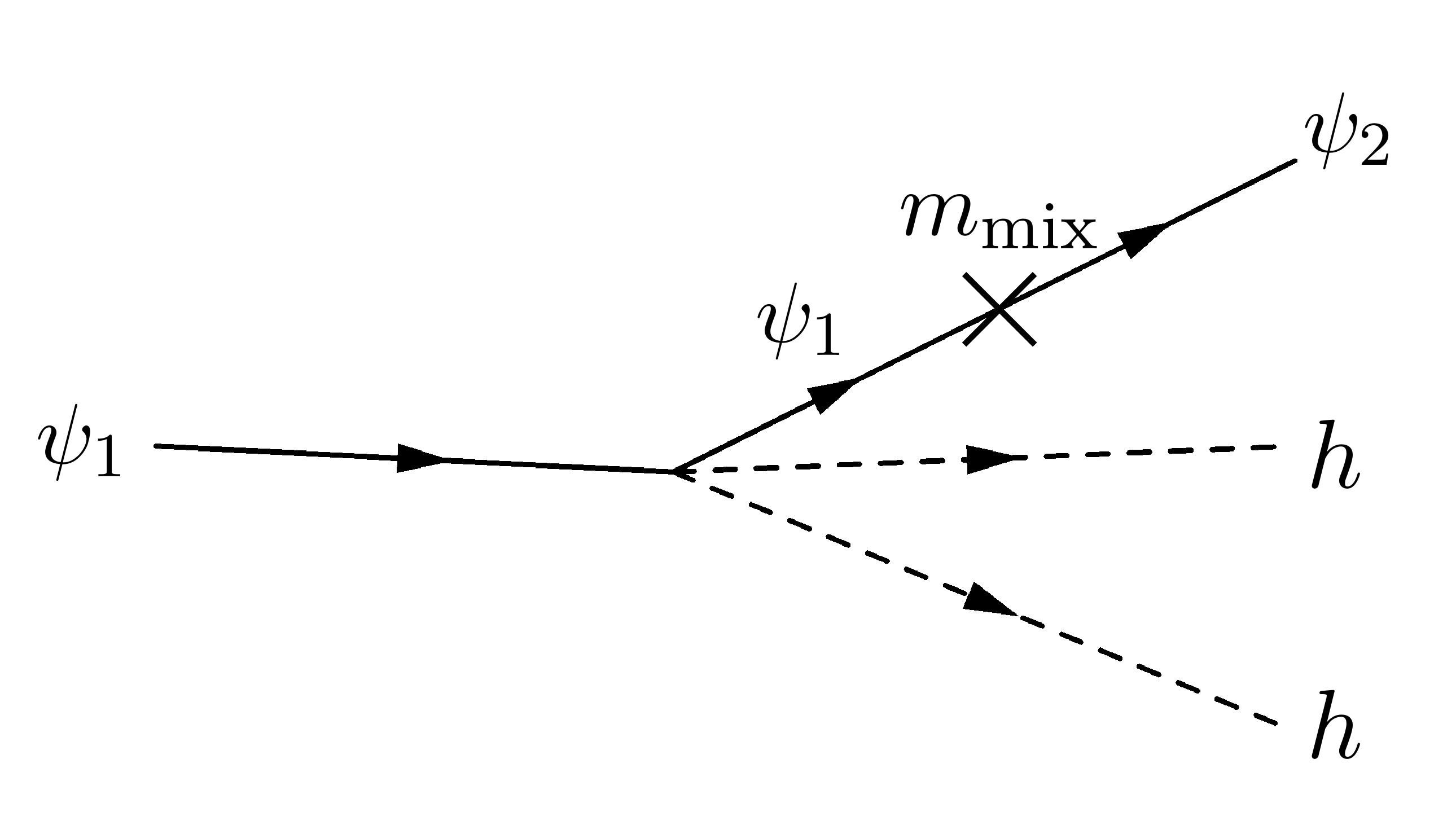}
  \label{fig:sub1}
\caption{If there is mass mixing, the term responsible for canceling quadratic divergences can give rise to decay channels. This signature is the unique indirect signature that one is led to in a bottom up approach to finding naturalness.}
\label{Fig: mixing}
\end{figure}

Additional interaction terms in the Lagrangian typically induce phenomenology of their own, independent of whether the quadratic divergences are canceled or not.  However there is a unique effect which can combine with quadratic divergences to yield detectable signals, mass mixing.
As shown in Fig.~\ref{Fig: mixing}, combining mass mixing with the bilinear interaction term given in Eq.~\ref{Eq: bilinear yet again} yields a decay channel.  Observation of this new decay channel gives indirect evidence that the new particle observed is involved with the Higgs quadratic divergences.

While non-renormalizable terms are necessarily UV dependent, renormalizable terms are not.  e.g. the soft IR breaking of SUSY is expressed in terms of renormalizable soft terms whereas the UV supersymmetry in manifest in the marginal terms\footnote{In contrast, the shift symmetry in Little Higgs models constrains both renormalizable and nonrenormalizable terms.  The point being made is simply that the IR mass terms are potentially free of the constraints of the UV symmetry.}.  We can then use the renormalizable mass mixing term, that is potentially unconstrained by the UV symmetry, in combination with the bilinear term to give an observable decay channel.  It should be noted however that one could simply have started with a term $\psi_1 \psi_2 H H^\dagger$ in the Lagrangian which emulates this effect.  As such, finding this decay channel provides only indirect evidence for the cancelation of quadratic divergences rather than direct evidence.

$\psi_1$ and $\psi_2$ in Fig.~\ref{Fig: mixing} could both be new particles or one could be a SM particle.  They decay in cascades involving only Higgses and not the associated Ws or Zs that one expects from the familiar trilinear Yukawa interactions.  If the $\psi_{1,2}$ are charged under the SM gauge group, then they can be pair produced via gauge bosons.  They decay to a Higgs and either a SM particle, a CHAMP or a R-hadron.  If  the $\psi_{1,2}$ are not charged under the standard model, then additional assumptions are required for how they might be produced at the LHC.

If $\psi_2$ is a SM field, then $\psi_1$ has the same quantum numbers as a Standard model particle and we have a vector like fourth generation.  As an explicit example, assume that $\psi_1$ has the same quantum numbers as a right handed up-type quark.  Writing down all interactions up to dimension 5, we have
\bea
\label{Eq: mixing1}
\mathcal{L} = m^i \psi_1 u_i^c + m_{\psi_1} \psi_1 \psi_1^c + \lambda_1^i \psi_1^c H Q_i + \lambda_2^{ij} u_i^c H Q_j +  \frac{\lambda_3}{m_{\psi_1}} \psi_1^c \psi_1 HH^\dagger +  \frac{\lambda_4^i}{m_{\psi_1}} u_i^c \psi_1 HH^\dagger
\eea
where $i$ and $j$ are flavor indices.  $\psi_1$ has tree level mixing with the quarks.  Working in the small vev limit and diagonalizing the mass terms in the small vev limit yields
\bea
\label{Eq: after mixing}
\mathcal{L}_\text{mixing} = m_U U U^c + \lambda_U^i U^c H Q_i + \lambda_{SM}^{ij} u_i^c H Q_j + \frac{\lambda_{UU}}{m_U} U^c U H H^\dagger +\frac{ \lambda_{Uu}^i}{m_U} u^c_i U H H^\dagger
\eea
The mass eigenstate of $\psi_1$ is called $U$.  $\lambda_{UU}$ is related to the structure of the quadratic divergences and is not directly measurable.  Due to mixing effects it is related to $\lambda_{Uu}$.  In the case where $\lambda_3=0$ or $\lambda_4=0$, as in Little Higgs models, we notice that $\lambda_{Uu}$ and $\lambda_{UU}$ are directly related by mixing angles.

  \begin{figure}
\centering
  \includegraphics[width=4in]{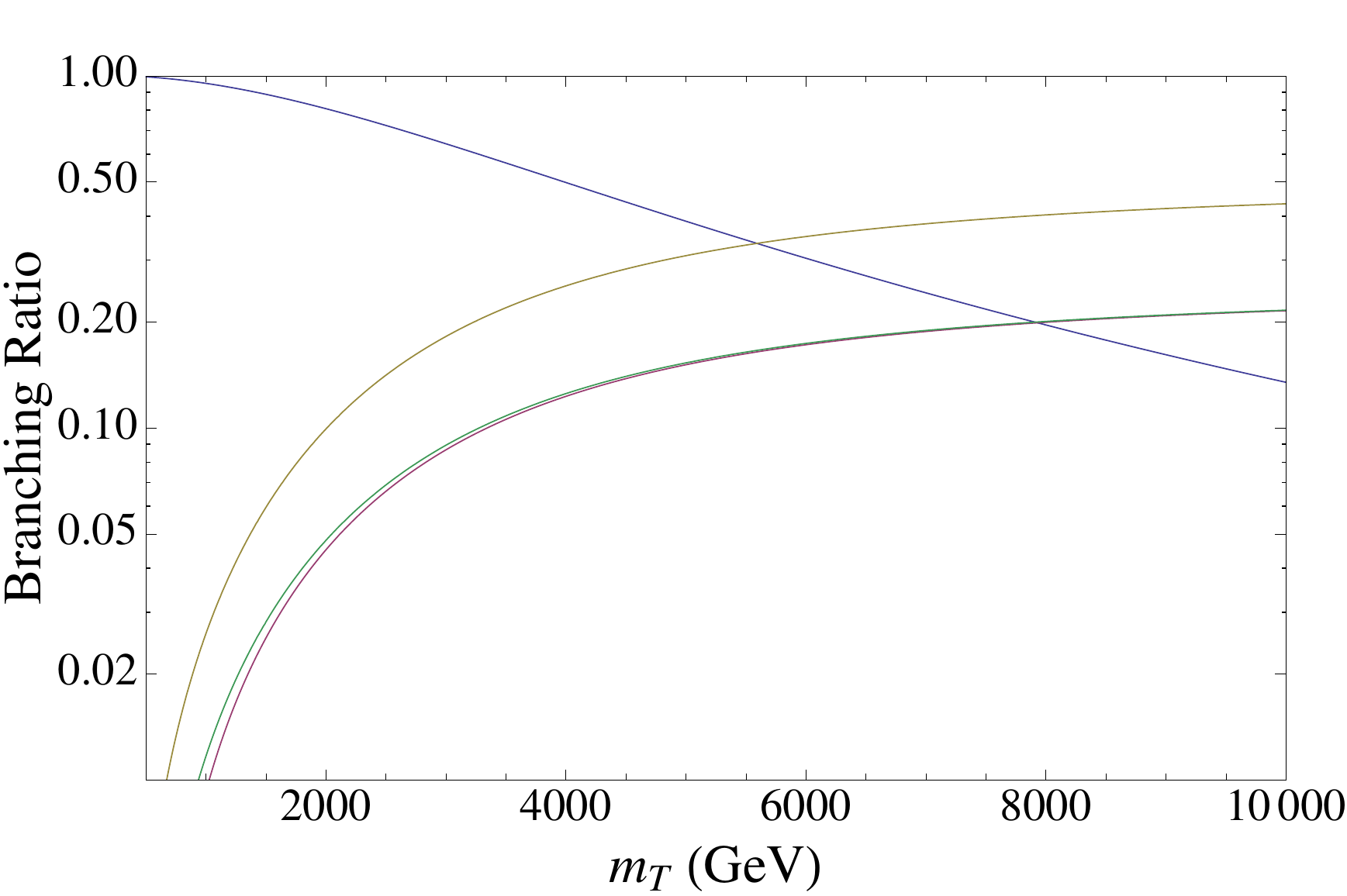}
  \label{fig:sub1}
\caption{The branching ratios into the four different decay channels $\Gamma(T\rightarrow h t)$, $\Gamma(T\rightarrow W^+ W^- t)$, $\Gamma(T \rightarrow h h t)$ and  $\Gamma(T \rightarrow Z Z t)$ with the colors blue, yellow, red and green respectively.  For masses above $m_X \sim 5$ TeV, the two body decay mode ceases to be the leading decay mode.}
\label{Fig: BranchingRatio}
\end{figure}

Observation of the decay channel $U \rightarrow u H H^\dagger$ gives strong indirect evidence that $\lambda_{UU} \ne 0$ and that the particle $U$ may be involved in canceling the quadratic divergences.  This evidence is indirect because it is always possible that $\lambda_3$ and $\lambda_4$ are chosen exactly such that $\lambda_{Uu} \ne 0$ while $\lambda_{UU}=0$. Generically, this miraculous cancelation does not occur. 

The new particle $U$ has decay channels resulting from $\lambda_{Uu}$ and $\lambda_U^i$. 
If $\lambda_U^i > \lambda_{Uu}^i\frac{v}{m_U}$, the decays from $\lambda_U^i$ are dominant, and the situation is simply that listed in Sec.~\ref{Sec: linear} where a term linear in H dominates the phenomenology.  Thus, we will focus on the case where $\lambda_{Uu}$ dominates over $\lambda_U^i$.  This limit can be realized in Little Higgs theories as shown in App.~\ref{Sec: Little Higgs}.

  Aside from the desire to categorize all models solving the naturalness problem, there are two other reasons why one might expect that the signatures associated with $\lambda_{Uu}$ dominate.  Generically, we expect $\mathcal{O}(1)$ mixing with no ad hoc cancelations so that we have $\lambda_U \sim \lambda_{SM1}$ and $\lambda_{UU} \sim \lambda_{Uu} \sim \frac{\lambda_{SM2}^2}{2 m_U}$, where $\lambda_{SM1}$ and $\lambda_{SM2}$ are the Yukawa couplings associated with the SM particle $\psi$ mixes with and the quadratic divergences cancelled by $\psi$ respectively.  In the case where $\psi$ is mixing with a light quark, but is canceling the top quadratic divergence, we get that $\lambda_U^i$ is suppressed by the light quark Yukawa so that $\lambda_{Uu}^i$ dominates.

Another reason why $\lambda_U^i$ terms may be suppressed is that $\lambda_U^i$ plays a role in flavor physics and contributes to flavor changing neutral currents~(FCNC).  In fourth generation and Little Higgs models, the typical assumption is that there is some ad hoc texture such that FCNCs are all avoided.  Analogously, we can assume that $\lambda_U^i$ is small due to flavor physics so that FCNCs are not an issue.  
$\lambda_U^i$ small also implies that the STU \cite{Peskin:1990zt} parameters are small.

$\lambda_{Uu}^i$ in Eq.~\ref{Eq: after mixing} can give four decay channels.  The decay channels are 
\bea
\label{Eq: decays}
U &\rightarrow& u_i + h \\
U &\rightarrow& u_i + h + h \\
U &\rightarrow& u_i + Z + Z  \\
U &\rightarrow& u_i + W^+ + W^-
\eea 
The three body decays can be the dominant or subdominant decay mode of $U$.  Their relative branching ratios are shown in Fig.~\ref{Fig: BranchingRatio}.  Unlike two body decays resulting from a Yukawa term, two body decays resulting from Eq.~\ref{Eq: bilinear yet again} involve only the Higgs and not the gauge bosons.  As shown in App.~\ref{Sec: goldstone}, this unique structure is important for maintaining the goldstone boson equivalence theorem. 

\begin{table}[t]
\centering
\begin{tabular}{|ccc||c|}
\hline
\multicolumn{2}{|c}{Quantum numbers}&SM particle&LHC final states
\\
$(SU(3),SU(2))_{U(1)}$&spin&that is mixed with&\\
\hline\hline
$(3,1)_{2/3},(3,1)_{-1/3},(\bar{3},2)_{-1/6}$&1/2&$u^c_{1,2}$,$d^c_{1,2}$,$Q_{1,2}$&$2q + H^\dagger H$ \\
$(3,1)_{-1/3}$&1/2&$d^c_3$&$2b + 2H^\dagger H$\\
$(3,1)_{2/3}$&1/2&$u^c_3$&$2t + 2H^\dagger H$ \\
$(\bar{3},2)_{-1/6}$&1/2&$Q_3$&$2 b/t + 2H^\dagger H$\\
$(1,1)_{-1}$&1/2&$e^c_{1,2}$&$l^+ l^- + 2H^\dagger H$ \\
$(1,1)_{-1}$&1/2&$e^c_{3}$&$\tau^+ \tau^- + 2H^\dagger H$ \\
$(1,2)_{1/2}$&1/2&$L^c_{1,2}$&$l^\pm/\nu +  l^\mp/\nu + 2H^\dagger H$ \\
$(1,2)_{1/2}$&1/2&$L^c_{3}$&$\tau^\pm/\nu +  \tau^\mp/\nu + 2H^\dagger H$\\
$(1,2)_{-1/2}$&0&$H$&$h/W/Z + h/W/Z + 2H^\dagger H$\\
\hline
\end{tabular}
\caption{The LHC signatures of various simplified models that require adding only a single new particle that mixes with a lighter state and cancels quadratic divergences.  The final state $H^\dagger H$ can stand for $h$, $W^+ W^-$, $Z^0 Z^0$, and $h h$.  The branching ratios to these four different decay channels are shown as a function of mass in Fig.~\ref{Fig: BranchingRatio}.  For low masses, the signature of canceling a quadratic divergence is the observation of a decay to a Higgs but not to W or Z bosons. }
\label{Tab: models}
\end{table}

At high $m_U$, the goldstone boson equivalence theorem implies that there will be only three body decays into pairs of Ws, Zs and hs with a ratio of 2:1:1. This structure is specific to the interaction term in Eq.~\ref{Eq: bilinear yet again}. Observation of three body decays into a pair of Higgs bosons and another particle is evidence that a new particle couples to the Higgs with a bilinear coupling.

Although the previous example illustrates the case of a new particle mixing only with an up-type quark, mixing with any other SM particle leads to similar conclusions, as shown in Table~\ref{Eq: decays}.  For masses that the LHC can probe, the two body decays will be seen first. The new particles will be pair produced through intermediate gauge bosons.

If there are two new particles which mix with each other rather than with the SM, the associated collider signatures will be cascade decays through
\bea
\psi_1 \rightarrow \psi_2 + H^\dagger H
\eea
where the final state $H^\dagger H$ can stand for $h$, $W^+ W^-$, $Z^0 Z^0$, and $h h$.  $\psi_2$ can result in missing $E_T$, a CHAMP, or a R-hadron depending on its quantum numbers.  In supersymmetric models, none of the new particles have the same quantum numbers as each other.  After electroweak symmetry breaking left and right handed sparticles can mix, but the effects of these mixings are too small to observe.

\subsection{Collider bounds on top quark, light quark and lepton partners}
\label{Sec: collider}
This subsection studies the sensitivity of the current ATLAS searches to the decays of quark and lepton partners of the form
\bea
\nonumber T &\rightarrow& t + h \\
 Q &\rightarrow& q + h \label{decays} \\
\nonumber L &\rightarrow& l + h
\eea
Pair produced quark or lepton partners give rise to collider signatures involving two Higgs bosons and two Standard Model particles.  Due to the wide variety of possible decay channels for a $125\GeV$ Higgs boson, there are many final states, often involving one or more lepton and/or b-jet.  Missing $E_T$ and leptons are needed to reduce the QCD background so that the identity of the Standard Model particle produced in association with the Higgs determines which search will be the most sensitive to the signal studied.

Each of the processes shown in Eq~\ref{decays} is studied separately and only signals corresponding to pair produced quark and lepton partners are investigated.  In what follows, hadronically decaying $\tau$ leptons are treated as jets. In the simulated samples, the leptonic partners $L$ decay to electrons and Higgs bosons and the light quark partners decay to up quarks and Higgs bosons. Details about the event generation and search validation are given in App.~\ref{Sec: MC}.

\subsubsection{Top quark partner}
A dedicated search for top partners decaying to a top quark and a Higgs boson already exists in ATLAS \cite{ATLAS:2013ima}. This search relies heavily on the b-jet multiplicity of the final states, the most sensitive signal region requiring exactly one lepton and four or more b-jets. The associated mass reach is about $800\GeV$ if $T\rightarrow t h$ is the only available decay channel.

\subsubsection{Electron partner}
\begin{table}
\centering
\begin{tabular}{lccc}
\toprule
& 0 b quarks &  2 b quarks & 4 b quarks \\
\midrule
2 leptons & 8\% & 33\% & 34\% \\
3 leptons & 6\% & 13\% & 0\% \\
4 leptons & 3\% & 3\% & 0\% \\
\bottomrule
\end{tabular}
\caption{Probabilities (in \%)  of getting a given number of leptons and b-quarks from decays of pair produced electron or muon partners. Detector effects and particle identification efficiencies are not included. \label{Tab: leptonEff}}
\end{table}

Decays of pair produced electron partners to an electron and a Higgs boson have particularly characteristic final states  with a large number of leptons and b-jets, as shown in Table~\ref{Tab: leptonEff}.  The high sensitivity of the ATLAS multilepton searches \cite{ATLAS:2013qla,ATLAS:2013rla} makes them particularly suited to such signals where about $20$\% of the final states have at least three leptons.

\begin{figure}
\centering
\begin{subfigure}{0.5\textwidth}
\includegraphics[width=\linewidth]{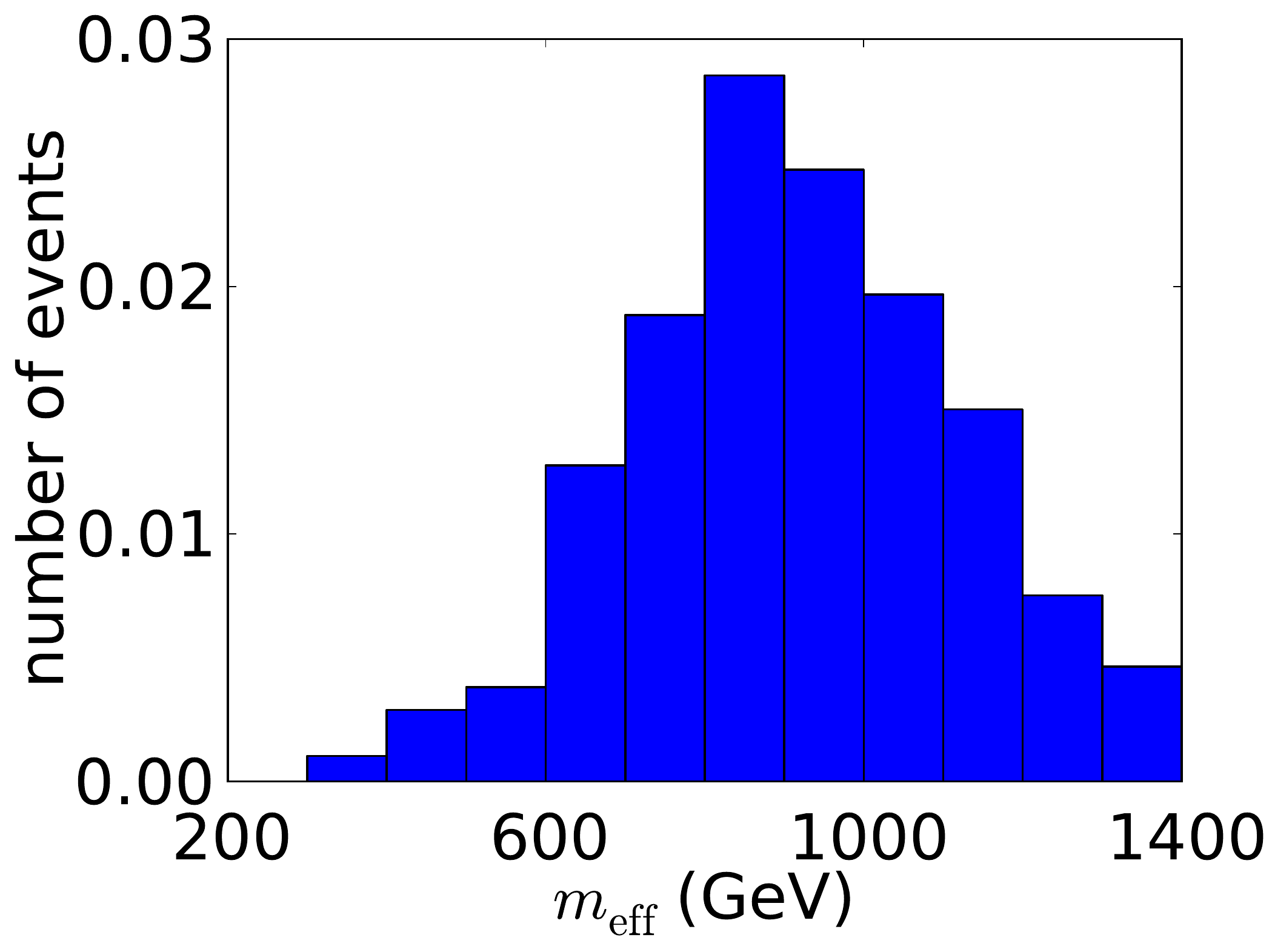}
\end{subfigure}
\caption{The effective mass distribution of a 500$\GeV$ electron partner with four leptons and no on-shell $Z$\label{meff}}
\label{Fig: distributions}
\end{figure}
\begin{figure}
\centering
\includegraphics[width=0.5\linewidth]{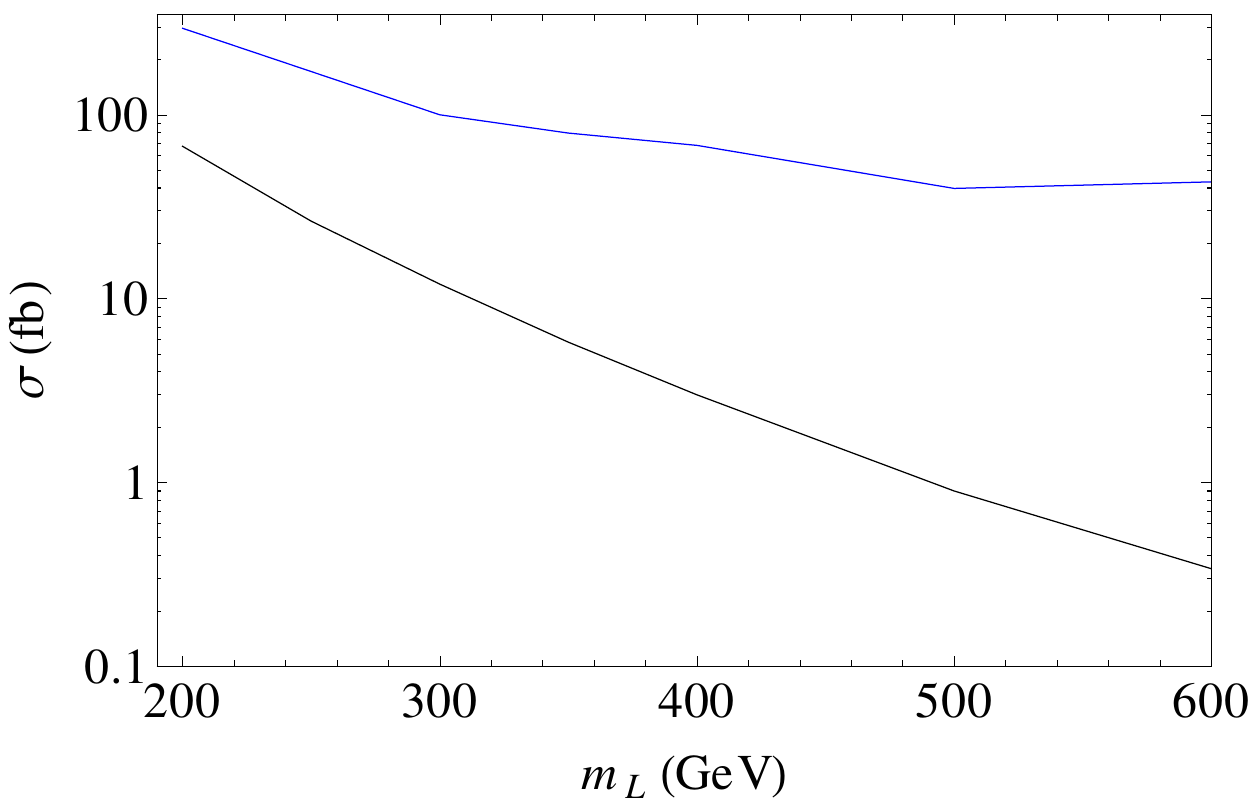}
\caption{$95$\% confidence limits on the pair production cross section of an electron partner which decays to a Higgs and an electron with a branching ratio of 1 (blue). The black line shows the LO pair production cross section of the electron partner\label{Fig: leptonbound}}
\end{figure}

Four lepton searches \cite{ATLAS:2013qla} have extremely low background rates and are sensitive to processes like the one studied.  In addition to the four lepton requirement, the effective mass $m_{eff}$ is required to be bigger than 600$\GeV$. $m_{eff}$ is defined as
\begin{align}
    m_{eff} = \sum_{l\in \mathrm{leptons}}p_{T_l}+\sum_{j\in \mathrm{jets},\text{ } p_{T_j}>30\GeV} p_{Tj}+\slashed{E_T} ,
\end{align}
As can be seen from the distributions in Fig~\ref{meff}, the efficiencies of these cuts are very high even for lepton partner masses as low as $m_L = 500\GeV$. For $m_L \ge 500\GeV$, the $95$\% confidence limit on the lepton partner production cross section has very low sensitivity to $m_L$ and is $\mathcal{O}(40) \fb$. For lower masses, signal regions with $50$ and $75\GeV$ missing $E_T$ cuts start being more competitive and the corresponding exclusion bounds are shown in Fig.~\ref{Fig: leptonbound}.  The electron partner pair production cross section has been computed at leading order (LO) using {\tt MadGraph 4.5.1}. As shown in Fig.~\ref{Fig: leptonbound}, getting reasonable sensitivity to $L\rightarrow l+H$ processes requires either going to higher energy and luminosity or designing a dedicated search.  

Three lepton searches \cite{ATLAS:2013rla} and searches requiring b-jets produced in association with two leptons \cite{atlas2013007,ATLAS:2013bma} could also have sensitivity to the final states shown in Table~\ref{Tab: leptonEff}. Most of the final states with three leptons are characterized by an opposite sign same flavor lepton pair (OSSF) with large invariant mass. The corresponding search region in the ATLAS 3 lepton search also requires $m_T>110\GeV$, with $m_T$ being the transverse mass of the lepton not belonging to the OSSF pair and the missing $E_T$. Since in most of the final states, the third lepton and the missing $E_T$ come from the decay of a $W$ or a $\tau$, this transverse mass requirement suppresses most of the signal. In the searches for two leptons produced in association with b-jets, the two leading leptons are required either to have the same sign or to come from the decay of an on-shell Z. The two leptons directly produced through the decay of the two lepton partners do not satisfy these requirements.

\subsubsection{Up quark partner}
Unlike top quark partners, no existing LHC search studies the decays of up quark partners to Higgs bosons and jets. The current exclusion bounds on quark partner masses are obtained by assuming a non-zero branching ratio to $W$ bosons plus jets \cite{Aad:2012bt}. The associated cuts heavily depend on the decay topology and cannot be transposed to the case studied here. Another search, specific to heavy bottom partners has been performed \cite{atlas2013051}, but it requires final states with exactly two leptons of the same sign, which can be obtained only through $B\rightarrow Wt$.

\begin{figure}
\centering
\includegraphics[width=0.5\linewidth]{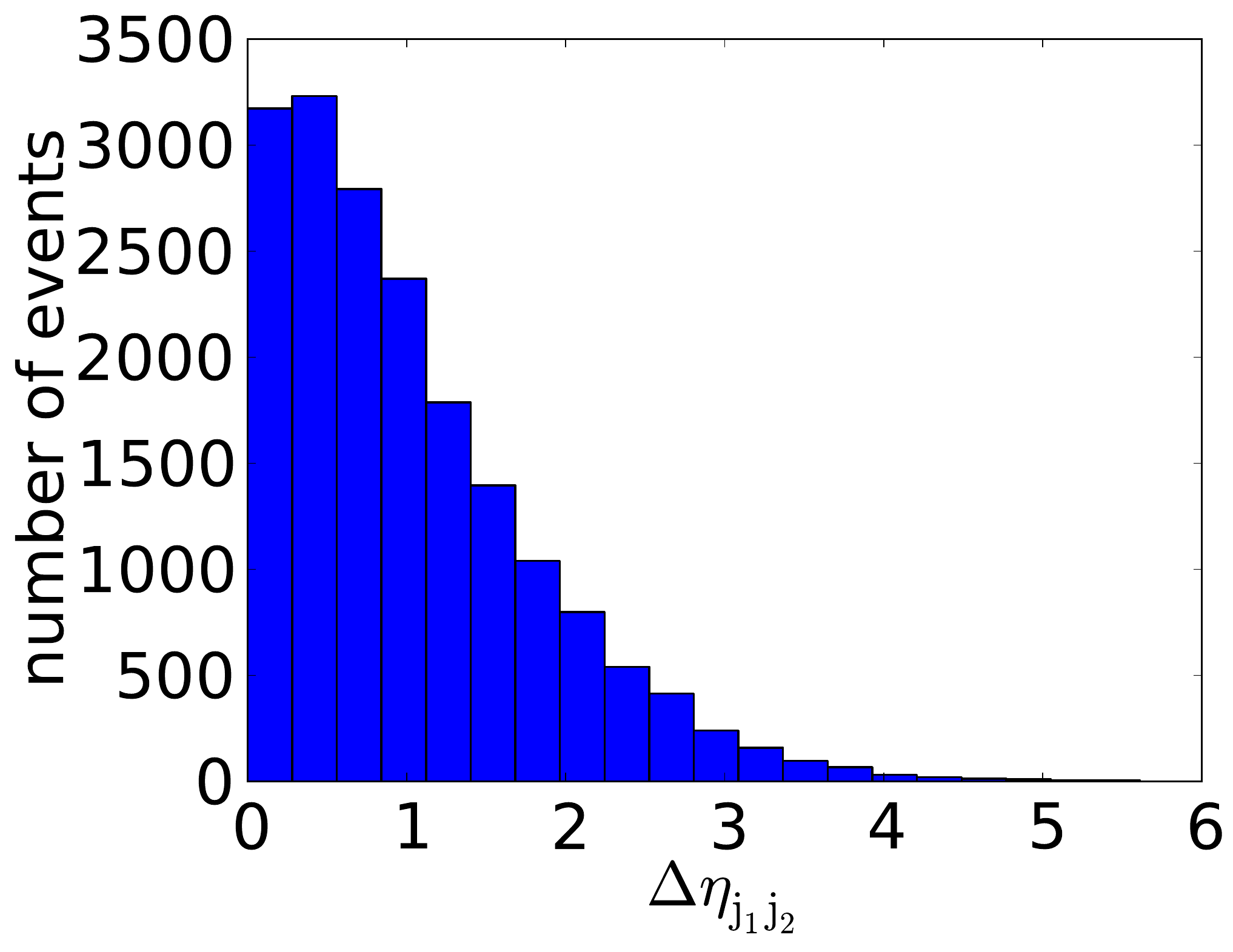}
\caption{Rapidity separation of the two leading jets for final states associated with pair produced up quark partners decaying to two Higgs bosons and two jets. The minimum rapidity separation required when looking for VBF topologies is of about $3$ units. \label{Fig: rap}}
\end{figure}

Decays of pair produced light quark partners lead to the production of two Higgs bosons and two jets.   Higgs searches, and especially searches looking for Higgs bosons produced through vector boson fusion (VBF), seem like they may be suited to this kind of signal. However, searches looking for Higgs bosons produced in association with two jets through VBF require a large rapidity gap. As can be seen from Fig.~\ref{Fig: rap}, this separation requirement proves particularly harmful for our signal. In all of the Higgs searches except the $H\rightarrow \gamma\gamma$ search, signal regions where the dominant single Higgs production mode is not VBF either require at most one jet or have exclusion bounds that are too loose at $125\GeV$.

The ATLAS $h\rightarrow\gamma \gamma$ search \cite{ATLAS:2012znl} sets reasonable exclusion bounds on the signal strength at $125\GeV$ and most of its signal regions do not have specific requirements on jets. This search is then the Higgs search providing the best sensitivity to our signal.  A $95$\% confidence bound on the $U\bar U$ production cross section can be derived by requiring that we do not produce more photons than the single SM Higgs does. Since the current exclusion bound on the signal strength for a $125\GeV$ Higgs is $\mu\sim 1.7$ and the single SM Higgs production cross section is about $20\pb$, it is not surprising that our exclusion bounds on the  $U\bar U$ production cross section are of the order of $10\pb$. Since no tight kinematic cuts are applied in the event preselection, these bounds are roughly independent of $m_U$. The associated mass reach is about $300\GeV$. 

Decays of the Higgs bosons through taus or vector bosons can lead to signals with 3 and 4 leptons. Unfortunately, due to the low branching ratio and the limited lepton identification efficiencies, the bounds found using the ATLAS 3 and 4 lepton searches are looser than the ones obtained using the $h\rightarrow\gamma \gamma$ search.

\section{Conclusion}
\label{Sec: conclusion}

This article studies model independent signatures of solutions to the naturalness problem. The Higgs 1-loop quadratic divergences can be cancelled by new particles interacting with the Higgs through either a Yukawa term or a quartic term. The first case is being probed by existing electroweakino or fourth generation searches. In the second case, if the new particle is a dark matter candidate, direct evidence of the existence of a bilinear quartic term could be around the corner for dark matter direct detection experiments. For scalar dark matter, a measurable correlation between  direct and indirect detection signals is also predicted.  More generally, if the particles canceling the quadratic divergences are charged under the SM gauge group, they could lead to modifications of the Higgs couplings to gauge bosons of up to $10$\%.  Alternatively, if the new particles obtain a vev, the Higgs couplings are suppressed.

In the case of a single new particle, we find that introducing mass mixing with a SM particle is the sole model independent method for obtaining an observable collider signature linked with quadratic divergences.  
 This mass mixing gives rise to two and three-body decay modes involving Higgs bosons. The two body decays arising from the quartic bilinear term dominate at sub-TeV masses and only involve Higgs bosons and not gauge bosons.
 
   Particles mixing with the top quark and decaying to a top quark and a Higgs are already constrained by the LHC to be heavier than $800\GeV$. Observing particles decaying to leptons or other quarks would require either going to higher energy and luminosity or designing dedicated searches. At the high luminosity LHC, new searches  could also isolate the three body decays of the heavy new particles to a pair of Higgses or gauge bosons and another SM particle.   A thorough study of the associated couplings and of possible correlations between the two and three-body decay rates would shed light on whether the decaying particle could cancel the Higgs quadratic divergences.

\section*{Acknowledgements}
The authors would like to thank Tim Cohen, Nathaniel Craig, Mariangela Lisanti and Jay Wacker for helpful discussions.  The authors would also like to thank Tim Cohen and Mariangela Listanti for comments on the draft.  SE is supported by a Stanford Graduate Fellowship.  AH is supported by the US DOE under contract number DE-SC0009988.

\section*{Note Added}

As the manuscript was being completed, Ref. \cite{Craig:2013xia} and  \cite{perelstein} which approached naturalness in a similar manner, appeared on the arXiv.

\appendix

\section{Goldstone Boson Equivalence Theorem}
\label{Sec: goldstone}

In this section, we clarify how the goldstone boson equivalence theorem work for quartic interactions.  In particular, we will study the system
\bea
\mathcal{L} \supset -m_T T T^c + \lambda H H^\dagger T u_3^c 
\eea
in the limit of large $m_T$ where the goldstone boson equivalence theorem holds.  In addition to the three body decays, there is the two body decay $T \rightarrow t h$ which is suppressed by a factor of $v/m_T$ relative to the three body decay.

\begin{figure}
\centering
\begin{tabular}{cc}
\begin{subfigure}{.5\textwidth}
  \centering
  \includegraphics[width=2.12in]{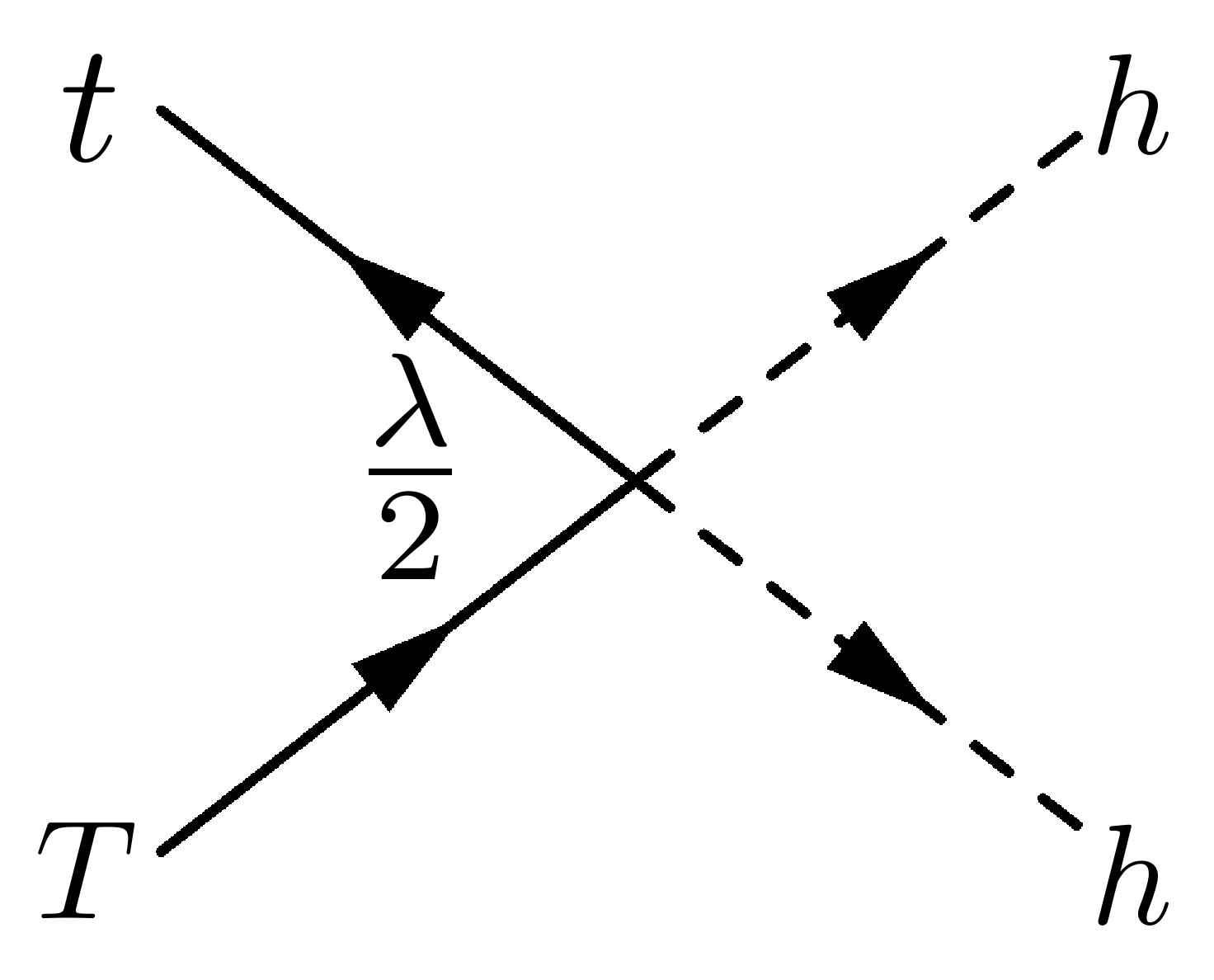}
\end{subfigure}
   &   \begin{subfigure}{.5\textwidth}
  \centering
\includegraphics[width=3.12in]{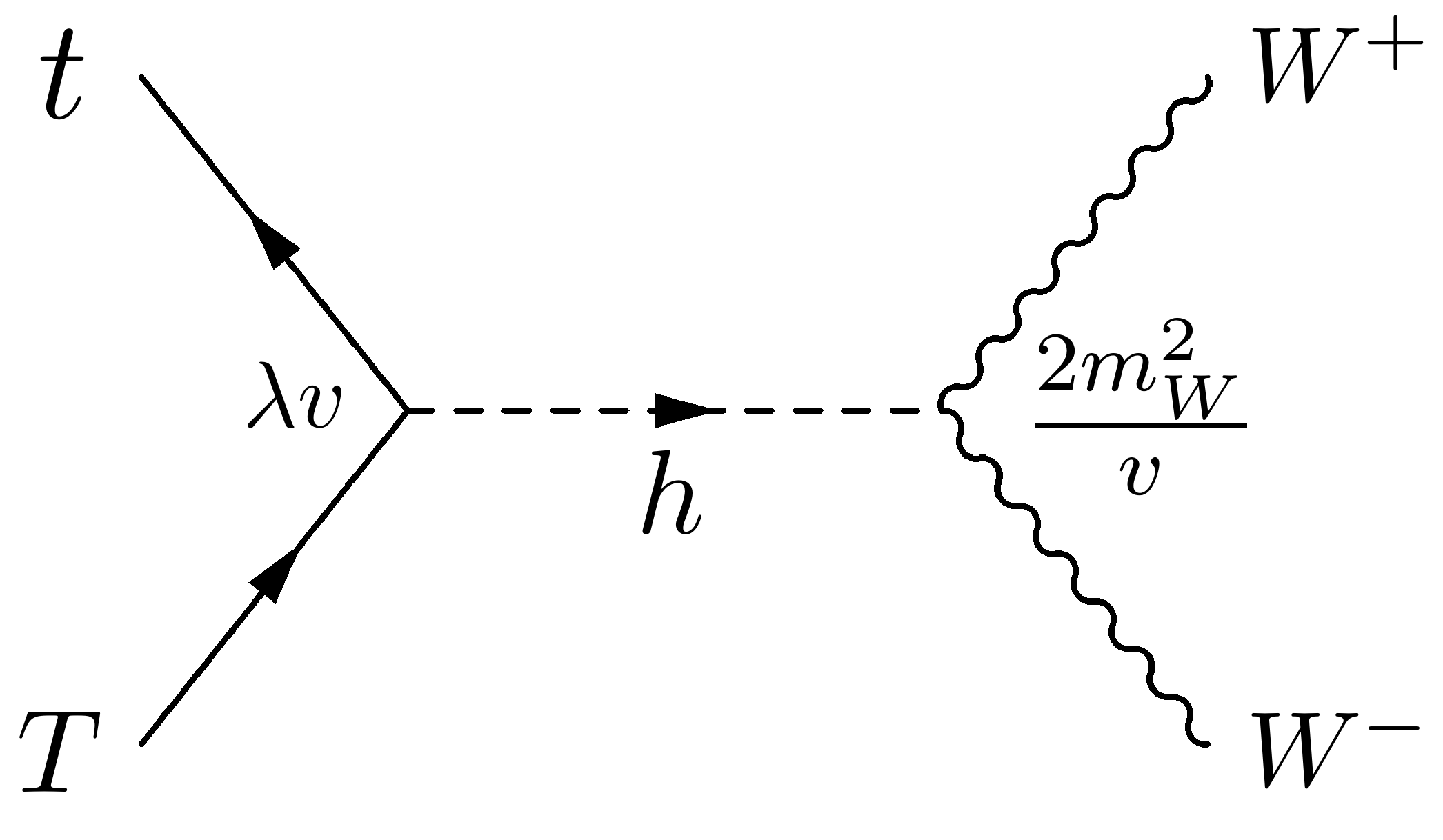}
\end{subfigure} 
\end{tabular}
\caption{}
\label{Fig: diagrams}
\end{figure}

The two three body decays that will be considered are $T \rightarrow t h h$ and $T \rightarrow t W^+ W^-$.  The two Feynmann diagrams are shown in Fig.~\ref{Fig: diagrams}.
Schematically the matrix elements are
\bea
|M(T \rightarrow t h h)|^2 &\sim& \frac{\lambda^2}{2} p_{T,\mu} p^{\mu}_t \\
\label{Eq: off shell}
|M(T \rightarrow t W^+ W^-)|^2 &\sim& 4 \lambda^2 m_W^4  p_{T,\mu} p^{\mu}_t \frac{1}{((p_T - p_t)^2-m_h^2)^2} \frac{(p_{W^+} \cdot p_{W^-})^2}{m_W^4} 
\eea
. The last term results from the polarization vectors of the massive gauge bosons which scale in the large energy limit as $\frac{k^\mu}{m_W}$.  

The usual expectation is that putting the Higgs on-shell would be the dominant contribution to Eq.~\ref{Eq: off shell}, i.e. the two body decay would be dominant.  In the large $m_T$ limit, this intuition is not valid because the large momentum flowing in the propagator is cancelled by the large $k^\mu$ in the polarization vectors of the gauge bosons.  Schematically, scalar and fermion final states contribute $k_\mu^{0,\frac{1}{2}}$ respectively so do not overcome the off-shell propagator suppression.

  Thus in the large $m_T$ limit, decays proceeding through an off-shell intermediate Higgs dominate.  This calculation shows that even though the two body decay of the Higgs is suppressed,  its existence is still important for maintaining the goldstone boson equivalence theorem.

\section{Little Higgs, Quadratic Divergences, and Three Body Decays}
\label{Sec: Little Higgs}

In Sec~\ref{Sec: bilinear indirect}, we considered bilinear couplings to the Higgs where the Yukawa term was suppressed.  In this section, we show how this limit can be reached in a toy Little Higgs model.  

Consider a toy Little Higgs models describing the spontaneous breaking of a $SU(3)$ down to an $SU(2)$ by a fundamental vev $f$ \cite{Perelstein:2003wd}.  The cut off scale at which new physics must enter is $4 \pi f$.  The breaking gives 5 pseudo-goldstone bosons, 4 of which can be made into the Higgs field and the other will be ignored as it plays no role in subsequent analysis.  The goldstones are combined into a non-linear sigma field $\Sigma$.
\bea
\Sigma = \exp{\left (\frac{i}{f}   \begin{pmatrix}
0 & H \\  H^\dagger  & 0
\end{pmatrix} \right )}  \begin{pmatrix}
0 \\  f
\end{pmatrix}
\eea

A pair of colored weyl fermions, $u'$ and $u'^c$, are added to cancel the top quadratic divergence by collective symmetry.  One of the weyl fermions combines with $Q_3$ to form a $SU(3)$ triplet $\chi = (Q_3 , u')$.  The lagrangian is then
\bea
\mathcal{L} \supset \lambda_1 u_3^c \Sigma \chi + \lambda_2 f u'^c u'
\eea
The leading $1/f$ terms in the lagrangian are then
\bea
\label{Eq: not diagonalized}
\mathcal{L} \supset f (\lambda_1 u_3^3 + \lambda_2 u'^c) u' - \lambda_1 u_3^c H Q_3 + \frac{\lambda_1}{2f} H H^\dagger u_3^c u'
\eea
After the Higgs obtains a vev, the last two terms of Eq.~\ref{Eq: not diagonalized} allow the top partner to decay via a higgs boson and a top.

\subsection{$v \ll f$ limit}

The $v \ll f$ limit is not always physical but it illustrates which term of the two terms in Eq.~\ref{Eq: not diagonalized} dominates the decay.  Diagonlizing the  mass terms, gives 
\bea
\mathcal{L} \supset \frac{\lambda_1 \lambda_2}{\sqrt{\lambda_1^2 + \lambda_2^2}} t^c_3 H Q_3 + \frac{\lambda_1^2}{\sqrt{\lambda_1^2 + \lambda_2^2}} T^c H Q_3 + \frac{\lambda_1^2}{2 m_T} H H^\dagger T^c T + \frac{\lambda_1 \lambda_2}{2 m_T} H H^\dagger t^c_3 T
\eea

Notice that the coupling constants associated with the three and four body interactions can be parametrically different.  Their ratio is
\bea
\label{Eq: ratio}
\frac{\lambda_2}{\lambda_1} \frac{\sqrt{\lambda_1^2 + \lambda_2^2}}{2} = \frac{\lambda_2^2}{2 y}
\eea
where we used that the top Yukawa is 
\bea
y = \frac{\lambda_1 \lambda_2}{\sqrt{\lambda_1^2 + \lambda_2^2}}
\eea
When comparing decay rates, this quantity gets squared and if large can overcome the smaller three-body phase space.  Depending on the value of $\lambda_2$, one can interpolate between the dominant decay being two body or three body.

If the Yukawa being generated is not the top Yukawa, but one of the smaller Yukawas then the ratio in Eq.~\ref{Eq: ratio} is naturally very large.  Thus, for non-top partners we expect the quartic interaction to dominate over the Yukawa interaction.

To further illustrate this model, consider top and bottom partners with various $\lambda_2$.  As before, we work in the limit where $v \ll f$.  We use the goldstone boson equivalence theorem to obtain the three body decay rates so that the ratio of their matrix elements squared are thus $2:1:1$.  Since the three body phase space is especially sensitive to the mass of the decay products, it is only at high masses that the decay widths approach this ratio as well.  

\begin{figure}
\centering
\begin{tabular}{cc}
\begin{subfigure}{.5\textwidth}
  \centering
  \includegraphics[width=3.12in]{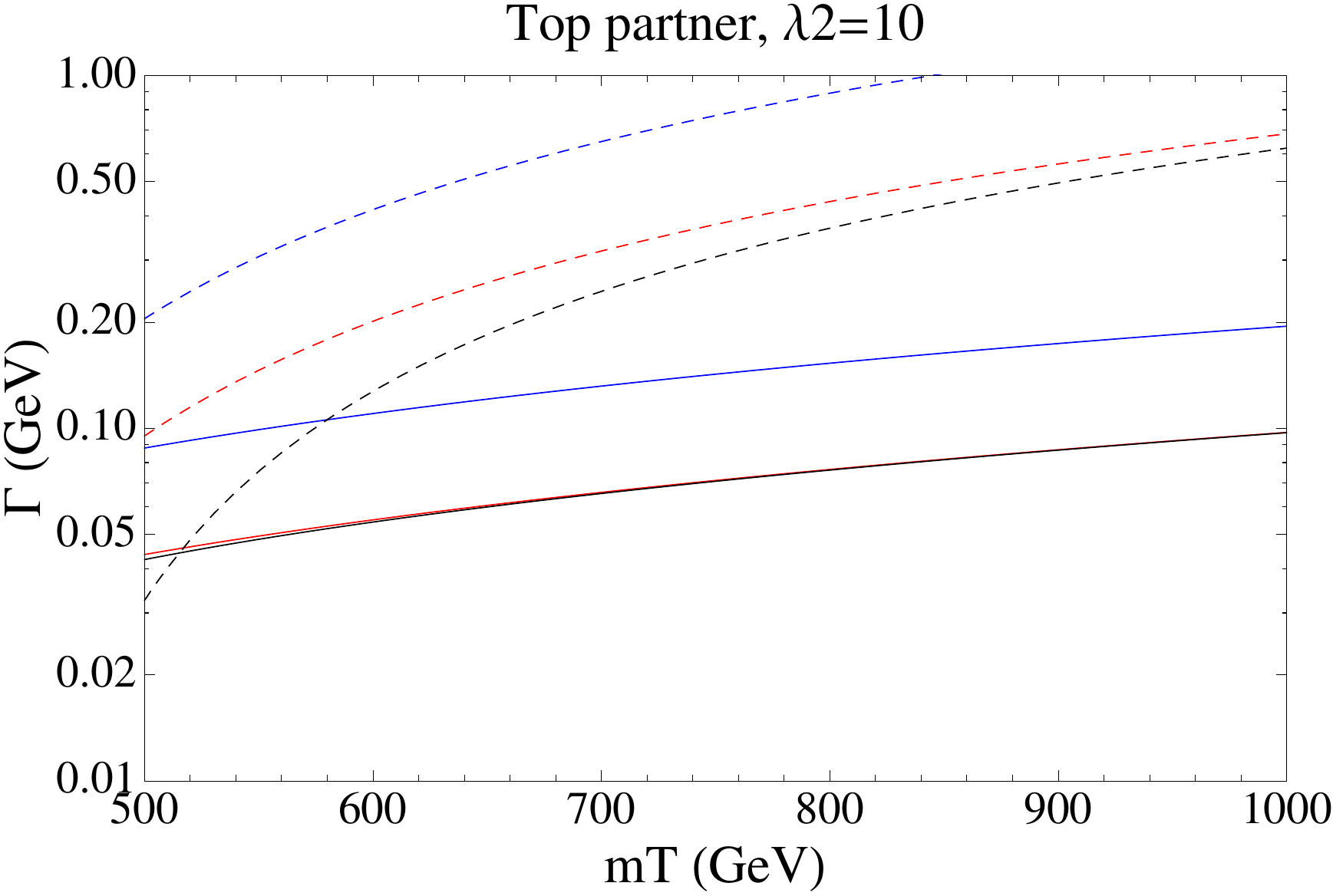}
  \label{fig:sub1}
\end{subfigure}
   &   \begin{subfigure}{.5\textwidth}
  \centering
\includegraphics[width=3.12in]{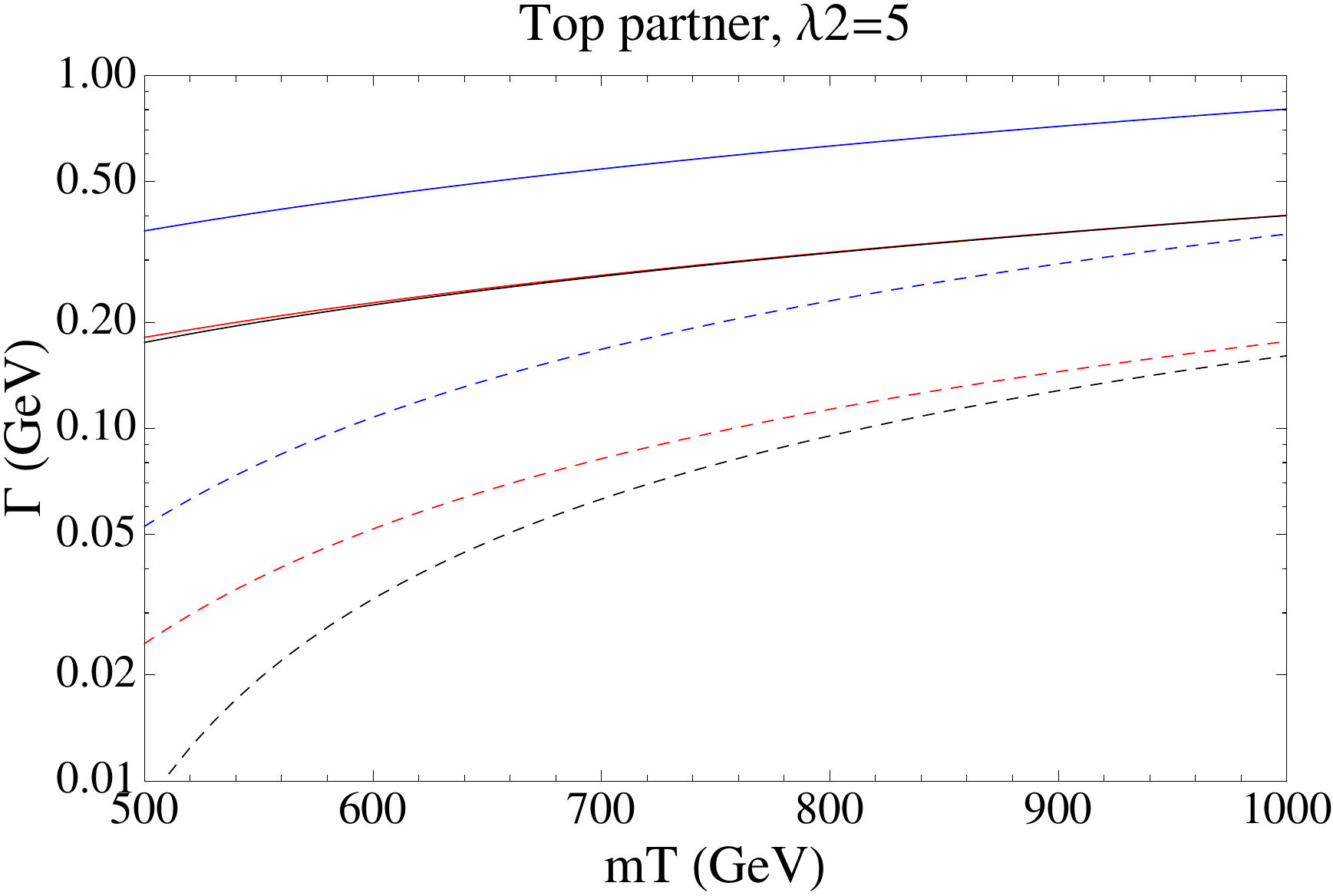}
  \label{fig:sub2}
\end{subfigure} \\
\begin{subfigure}{.5\textwidth}
  \centering
  \includegraphics[width=3.12in]{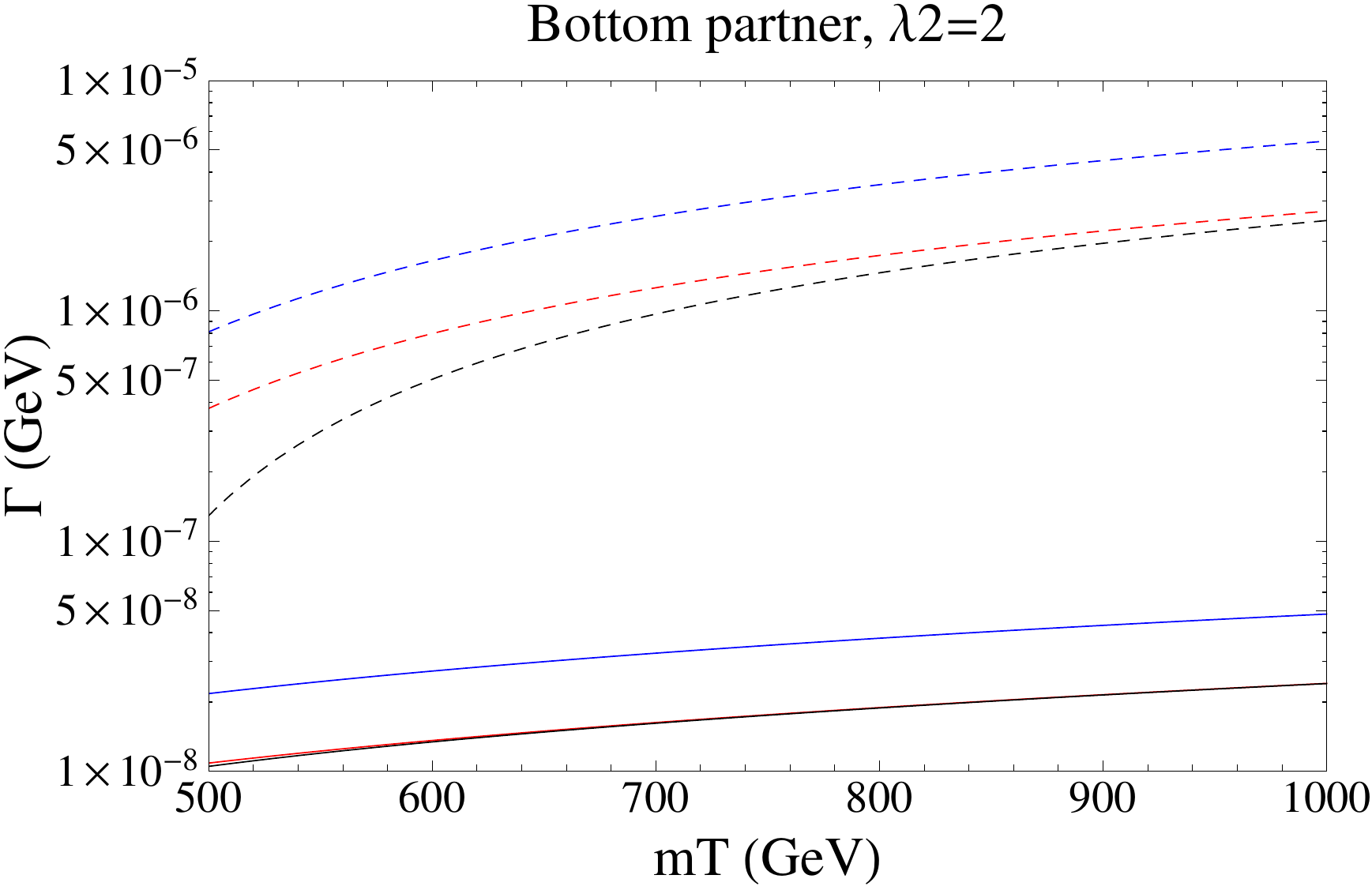}
  \label{fig:sub1}
\end{subfigure}
   &   \begin{subfigure}{.5\textwidth}
  \centering
\includegraphics[width=3.12in]{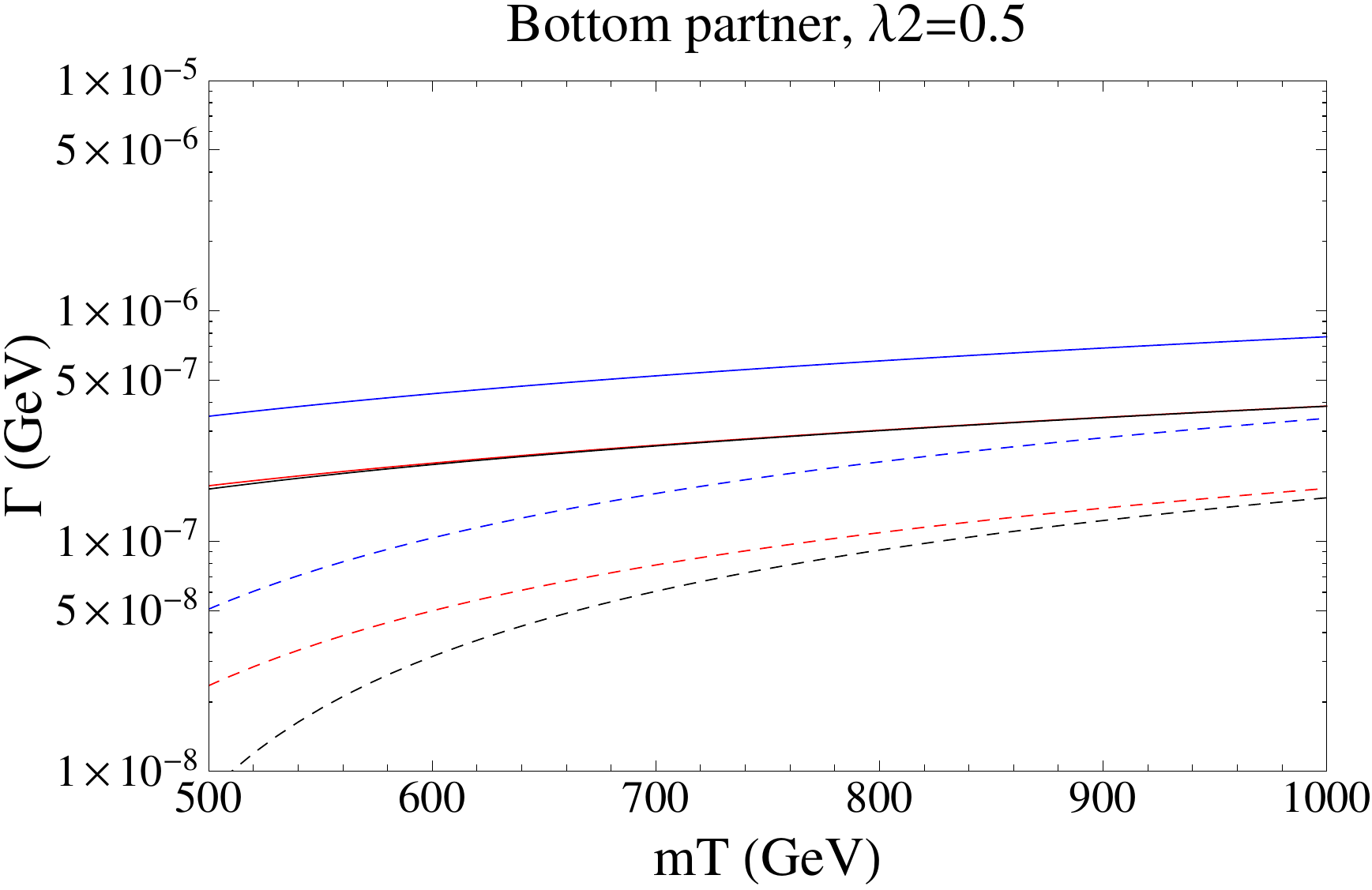}
  \label{fig:sub2}
\end{subfigure}
\end{tabular}
\caption{The dependence of the partial decay widths on the mass of the top/bottom partner and $\lambda_2$ in the $v \ll f$ limit.  The solid blue/red/black lines show $\Gamma(T\rightarrow W b)$, $\Gamma(T\rightarrow Z t)$ and $\Gamma(T\rightarrow h t)$ respectively.  The dashed blue/red/black lines show $\Gamma(T\rightarrow W^+ W^- t)$, $\Gamma(T\rightarrow Z Z t)$ and $\Gamma(T\rightarrow h h t)$ respectively.  The goldstone equivalence theorem was used to calculate the three body decays.  In the case of a top partner, it is hard for the 3 body decay to be parametrically larger than the 2 body decay, but for a bottom partner it happens quite readily.}
\label{Fig: top}
\end{figure}

In Fig.~\ref{Fig: top}, we see that one can interpolate between making the 3 body decay dominant and the 2 body decay dominant by varying $\lambda_2$.  For a top quark partner, it is difficult to make the 3 body decay parametrically dominant while for a bottom quark partner, the three body decay is expected to be the dominant decay mode.

\subsection{$v \sim f$}

In the previous subsection, the critical assumption of $v \ll f$ was made.  The assumption is invalidated  at low masses because if $m_T \sim 500$ GeV and $\lambda_2 \sim 10$ then $f \sim 50$ GeV.  The reason why $v \ne 0$ leads to such strong constraints is that $H H^\dagger u_3^c u'$ always leads to a two body decay when H obtains a vev.  One can estimate what the maximal relative branching ratios are as a function of $m_T$.  Assuming massless decay products, one obtains the relationship
\bea
\frac{\Gamma(T\rightarrow h t)}{\Gamma(T\rightarrow h h t)} \sim \frac{m_T^2}{96 \pi^2 v^2}
\eea
This ratio shows that for large $\lambda_2$, the three body decay to two Higgses dominates over the two body decay around 8 TeV.  The exact Little Higgs computation confirms this expectation.

\begin{figure}
\centering
\begin{tabular}{cc}
\begin{subfigure}{.5\textwidth}
  \centering
  \includegraphics[width=3.12in]{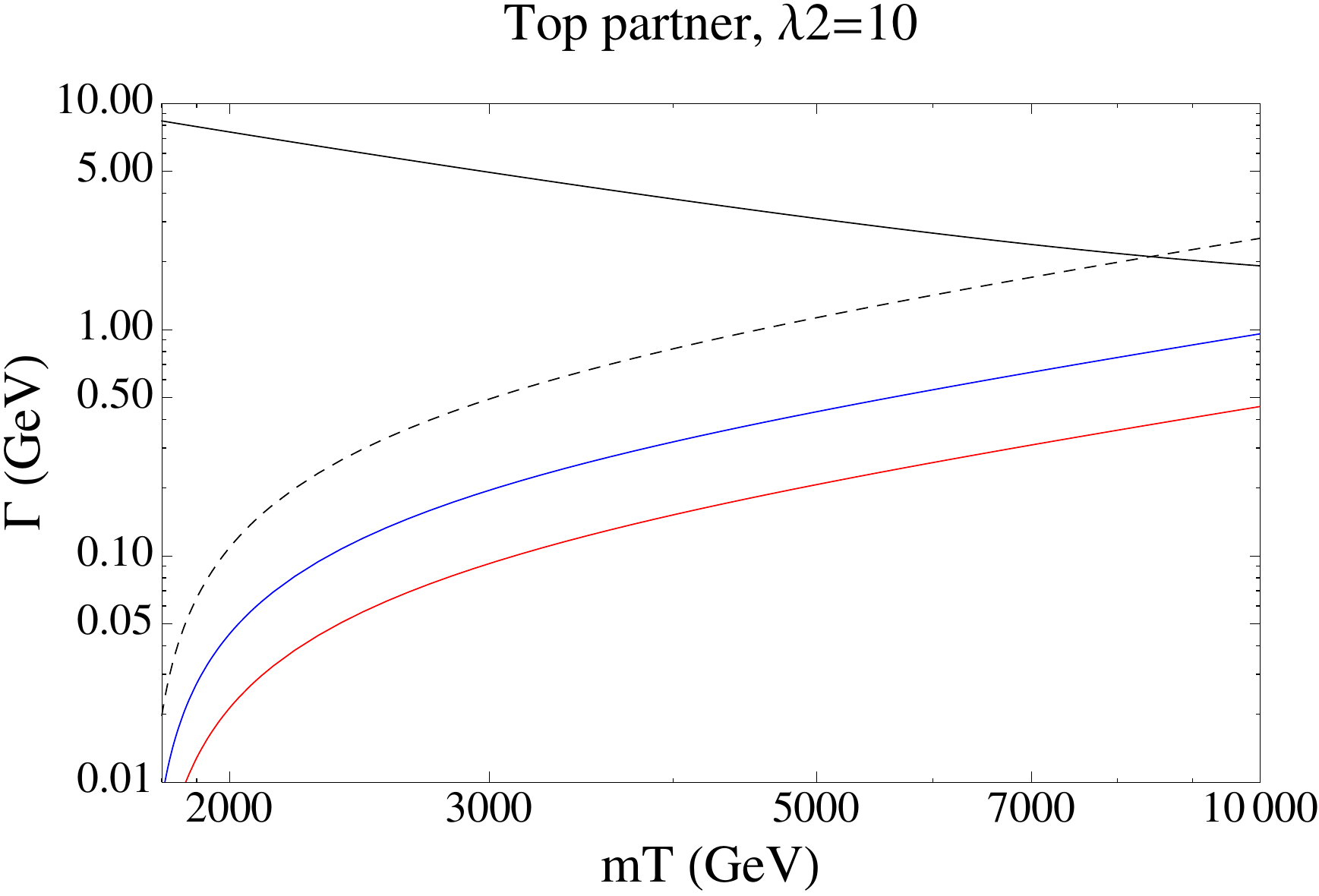}
  \label{fig:sub1}
\end{subfigure}
   &   \begin{subfigure}{.5\textwidth}
  \centering
\includegraphics[width=3.12in]{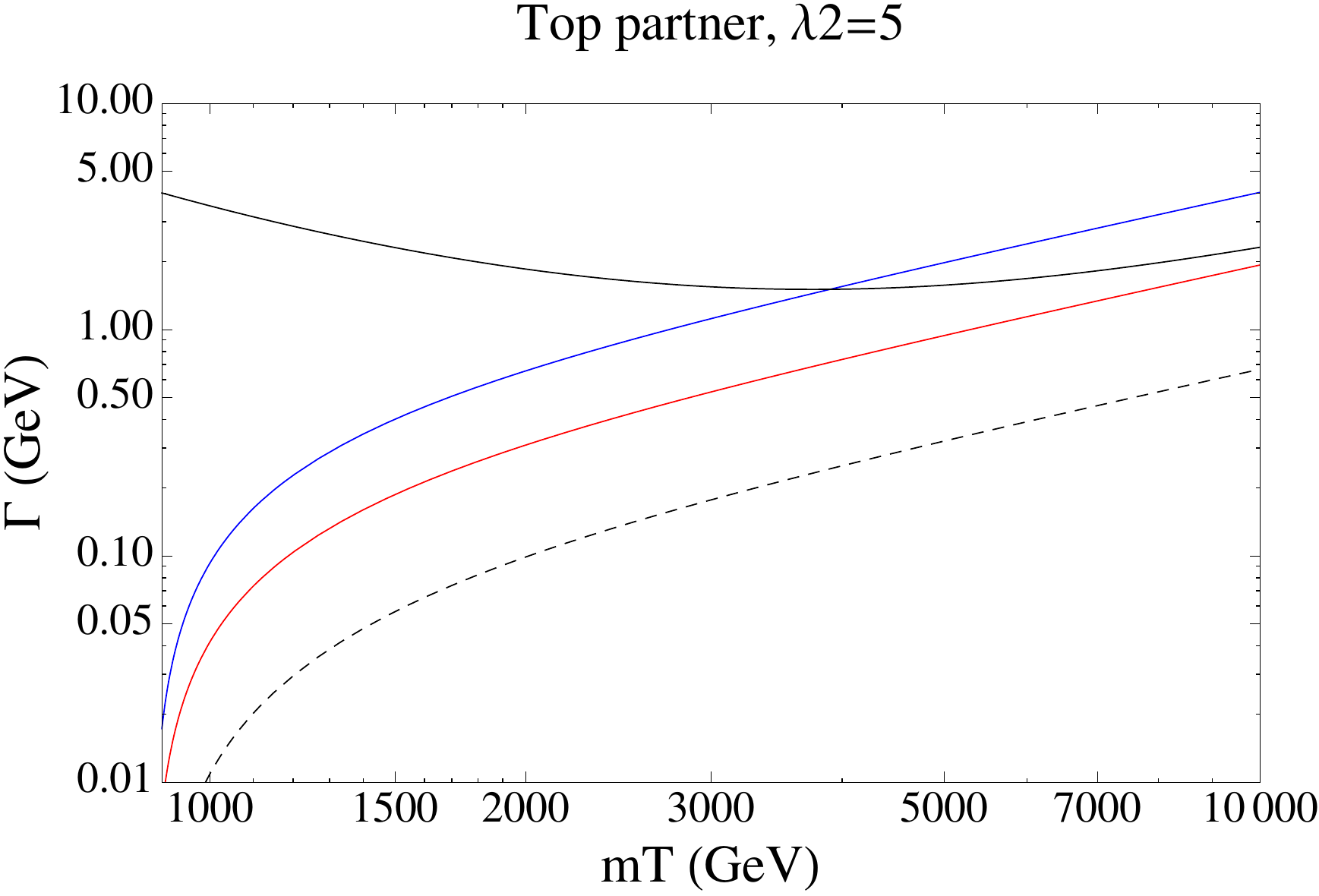}
  \label{fig:sub2}
\end{subfigure} 
\end{tabular}
\caption{The dependence of the partial decay widths on the mass of the top partner and $\lambda_2$ when $v \sim f$.  The solid blue/red/black lines show $\Gamma(T\rightarrow W b)$, $\Gamma(T\rightarrow Z t)$ and $\Gamma(T\rightarrow h t)$ respectively.  The dashed black line shows $\Gamma(T\rightarrow h h t)$.  Even though the three body is not dominant except for large masses and $\lambda_2$, it can be a relevant decay channel even at lower masses and $\lambda_2$.}
\label{Fig: top2}
\end{figure}

  Fig.~\ref{Fig: top2} shows the differential cross section as a function of $m_T$.  Except for rather large masses, it is impossible to make a three body decay dominate over all of the two body decays.  In the large $\lambda_2$ limit, the decay to bW and tZ are suppressed so that the only competing two body decay channel is $T \rightarrow t h$.  This suppression is an indication that what is causing that particular two body decay channel is in fact the quartic coupling rather than the Yukawa interaction.  This effect is simply what was discussed in Sec.~\ref{Sec: bilinear indirect}.

\section{Details of Monte Carlo Generation}
\label{Sec: MC}

Events were generated using {\tt MadGraph 4.5.1} \cite{Alwall:2011uj} and showered with {\tt Pythia 6.4} \cite{Sjostrand:2006za}. After showering, all hadron-level events are passed to the {\tt PGS 4} detector simulation, which parameterizes the detector response. The detector parameters used are those of the default ATLAS {\tt PGS} card except for jets, which are clustered using the anti-$k_t$ algorithm with a radius of $0.4$.
Heavy flavor jets are tagged using the $p_T$ dependent efficiencies found in \cite{CMS:2011cra}. In order to validate the ATLAS three and four lepton searches, light leptons identified by {\tt PGS} are tagged using the $p_T$ and $\eta$ dependent tight electron and muon identification efficiencies of respectively \cite{Aad:2011mk} and \cite{talk}. Hadronic taus are treated as jets.

For most SUSY searches, ATLAS gives the cut flow associated to a few benchmark models, which can then be used to validate our results. The results of this validation process for the ATLAS multilepton searches studied in Sec~\ref{Sec: collider} are shown in Table~\ref{Tab: 4leptons} and \ref{Tab: 3leptons}.

\begin{table}
\centering
\begin{tabular}{ccc}
\multicolumn{3}{c}{4 lepton search}\\
\hline
\hline
\multicolumn{3}{c}{$\chi_3^0\chi_2^0\rightarrow l^+l^-l^+l^-\chi_1^0\chi_1^0$}\\
\hline
\multicolumn{3}{c}{$m_{\chi_3^0}=310\GeV$ $m_{\chi_2^0}=305\GeV$}\\
\multicolumn{3}{c}{$m_{\chi_1^0}=230\GeV$}\\
\multicolumn{3}{c}{SR0noZa}\\
\hline
&Simulated & Expected\\
Lepton multiplicity & 55 & 77\\
Z veto & 45 & 38.6\\
missing $E_T$ cut & 27 & 25.6
\end{tabular}
\caption{Cut flow for the signal region SR0noZa of the ATLAS $20.7\ifb$ 4 lepton search \cite{ATLAS:2013qla}. The benchmark model tested here is $\chi_3^0\chi_2^0\rightarrow l^+l^-l^+l^-\chi_1^0\chi_1^0$. The decay of the neutralinos to leptons and LSP occurs through a slepton of mass $265\GeV$. For each step of the event selection, the number of events obtained using our simulated samples and our code is given on the left and the number of events given by ATLAS is shown on the right. \label{Tab: 4leptons}}
\end{table}
\begin{table}
\centering
\begin{tabular}{ccc|cc}
\multicolumn{5}{c}{3 lepton search}\\
\hline
\hline
\multicolumn{5}{c}{$\chi_1^\pm\chi_2^0\rightarrow W^\pm Z\chi_1^0\chi_1^0$}\\
\hline
\multicolumn{1}{c}{}&\multicolumn{2}{c}{$m_{\chi_1^\pm}=m_{\chi_2^0}=150\GeV$} & \multicolumn{2}{c}{$m_{\chi_1^\pm}=m_{\chi_2^0}=250\GeV$}\\
\multicolumn{1}{c}{}&\multicolumn{2}{c}{$m_{\chi_1^0}=75\GeV$} & \multicolumn{2}{c}{$m_{\chi_1^0}=0\GeV$}\\
\multicolumn{1}{c}{}&\multicolumn{2}{c}{SRnoZb} & \multicolumn{2}{c}{SRZc}\\
\hline
&Simulated & Expected & Simulated & Expected\\
Lepton multiplicity &197 & 227.3 & 36 & 40\\
OSSF requirement & 196 & 226.5 & 35 & 39.7\\
b veto & 196 & 211 & 35 & 36.4\\
Z veto/request & 182 & 196.6 & 32 & 34.4\\
missing $E_T$ & 38 & 53.8 & 16.5 & 17.7\\
$m_{\text{OSSF}}$ & 21 & 27.1 & 10.7 & 12.0\\
SRnoZc veto & 18 & 26.3 & -- & --
\end{tabular}
\caption{Cut flow for the signal region SR0noZa of the ATLAS $20.7\ifb$ 3 lepton search \cite{ATLAS:2013rla}. The benchmark model tested here is $\chi_1^\pm\chi_2^0\rightarrow W^\pm Z\chi_1^0\chi_1^0$. For each step of the event selection, the number of events obtained using our simulated samples and our code is given on the left and the number of events given by ATLAS is shown on the right.\label{Tab: 3leptons}}
\end{table}

\bibliographystyle{JHEP}

\begin{thebibliography}{10}

\bibitem{Chatrchyan:2012tx} 
  S.~Chatrchyan {\it et al.}  [CMS Collaboration],
  Phys.\ Lett.\ B {\bf 710}, 26 (2012)
  [arXiv:1202.1488 [hep-ex]].

\bibitem{ATLAS:2012ae} 
  G.~Aad {\it et al.}  [ATLAS Collaboration],
  Phys.\ Lett.\ B {\bf 710}, 49 (2012)
  [arXiv:1202.1408 [hep-ex]].

\bibitem{ArkaniHamed:1998rs} 
  N.~Arkani-Hamed, S.~Dimopoulos and G.~R.~Dvali,
  Phys.\ Lett.\ B {\bf 429}, 263 (1998)
  [hep-ph/9803315].
    
\bibitem{Randall:1999ee} 
  L.~Randall and R.~Sundrum,
  Phys.\ Rev.\ Lett.\  {\bf 83}, 3370 (1999)
  [hep-ph/9905221].

\bibitem{ArkaniHamed:2001nc} 
  N.~Arkani-Hamed, A.~G.~Cohen and H.~Georgi,
  Phys.\ Lett.\ B {\bf 513}, 232 (2001)
  [hep-ph/0105239].
 
\bibitem{ArkaniHamed:2002qy}
  N.~Arkani-Hamed, A.~G.~Cohen, E.~Katz and A.~E.~Nelson,
  JHEP {\bf 0207} (2002) 034
  [hep-ph/0206021].

\bibitem{Chacko:2005pe} 
  Z.~Chacko, H.~-S.~Goh and R.~Harnik,
  Phys.\ Rev.\ Lett.\  {\bf 96}, 231802 (2006)
  [hep-ph/0506256].

\bibitem{Burdman:2006tz} 
  G.~Burdman, Z.~Chacko, H.~-S.~Goh and R.~Harnik,
  JHEP {\bf 0702}, 009 (2007)
  [hep-ph/0609152].

\bibitem{Carmi:2012yp} 
  D.~Carmi, A.~Falkowski, E.~Kuflik and T.~Volansky,
  JHEP {\bf 1207}, 136 (2012)
  [arXiv:1202.3144 [hep-ph]].
  

\bibitem{Hall:2011aa} 
  L.~J.~Hall, D.~Pinner and J.~T.~Ruderman,
  JHEP {\bf 1204}, 131 (2012)
  [arXiv:1112.2703 [hep-ph]].

\bibitem{Brust:2011tb} 
  C.~Brust, A.~Katz, S.~Lawrence and R.~Sundrum,
  JHEP {\bf 1203}, 103 (2012)
  [arXiv:1110.6670 [hep-ph]].

\bibitem{Papucci:2011wy} 
  M.~Papucci, J.~T.~Ruderman and A.~Weiler,
  JHEP {\bf 1209}, 035 (2012)
  [arXiv:1110.6926 [hep-ph]].

\bibitem{Kats:2011qh} 
  Y.~Kats, P.~Meade, M.~Reece and D.~Shih,
  JHEP {\bf 1202}, 115 (2012)
  [arXiv:1110.6444 [hep-ph]].
 
\bibitem{Perelstein:2003wd} 
  M.~Perelstein, M.~E.~Peskin and A.~Pierce,
  Phys.\ Rev.\ D {\bf 69}, 075002 (2004)
  [hep-ph/0310039].
 
\bibitem{Berger:2012ec} 
  J.~Berger, J.~Hubisz and M.~Perelstein,
  JHEP {\bf 1207}, 016 (2012)
  [arXiv:1205.0013 [hep-ph]].

\bibitem{Fajfer:2013wca} 
  S.~Fajfer, A.~Greljo, J.~F.~Kamenik and I.~Mustac,
  arXiv:1304.4219 [hep-ph].
  
  
\bibitem{Alves:2011wf} 
  D.~Alves {\it et al.}  [LHC New Physics Working Group Collaboration],
  J.\ Phys.\ G {\bf 39}, 105005 (2012)
  [arXiv:1105.2838 [hep-ph]].

\bibitem{CMS:2012pva} 
  [CMS Collaboration],
  CMS-PAS-EXO-11-098.
  
\bibitem{Nektarijevic:2012zf} 
  S.~Nektarijevic [ATLAS Collaboration],
  PoS HQL {\bf 2012}, 070 (2012).

\bibitem{Frampton:1999xi} 
  P.~H.~Frampton, P.~Q.~Hung and M.~Sher,
  Phys.\ Rept.\  {\bf 330}, 263 (2000)
  [hep-ph/9903387].
  
\bibitem{Holdom:2009rf} 
  B.~Holdom, W.~S.~Hou, T.~Hurth, M.~L.~Mangano, S.~Sultansoy and G.~Unel,
  PMC Phys.\ A {\bf 3}, 4 (2009)
  [arXiv:0904.4698 [hep-ph]].
  
\bibitem{Aad:2011yf} 
  G.~Aad {\it et al.}  [ATLAS Collaboration],
  Phys.\ Lett.\ B {\bf 701}, 1 (2011)
  [arXiv:1103.1984 [hep-ex]].
  
\bibitem{Aad:2012pra} 
  G.~Aad {\it et al.}  [ATLAS Collaboration],
  Phys.\ Lett.\ B {\bf 720}, 277 (2013)
  [arXiv:1211.1597 [hep-ex]].

\bibitem{Farrar:1978xj} 
  G.~R.~Farrar and P.~Fayet,
  Phys.\ Lett.\ B {\bf 76}, 575 (1978).

\bibitem{Chatrchyan:2013oca} 
  S.~Chatrchyan {\it et al.}  [CMS Collaboration],
  arXiv:1305.0491 [hep-ex].

\bibitem{Achard:2001qw} 
  P.~Achard {\it et al.}  [L3 Collaboration],
  Phys.\ Lett.\ B {\bf 517}, 75 (2001)
  [hep-ex/0107015].
  
\bibitem{Aad:2011mb} 
  G.~Aad {\it et al.}  [ATLAS Collaboration],
  Phys.\ Lett.\ B {\bf 698}, 353 (2011)
  [arXiv:1102.0459 [hep-ex]].

\bibitem{ATLAS:2013sla} 
  [ATLAS Collaboration],
  ATLAS-CONF-2013-034.

\bibitem{CMS:yva} 
  [CMS Collaboration],
  CMS-PAS-HIG-13-005.

\bibitem{ATLAS:2013oma} 
  [ATLAS Collaboration],
  ATLAS-CONF-2013-012.

\bibitem{CMS:photons} 
  [CMS Collaboration],
  CMS-PAS-HIG-13-001.


\bibitem{Aprile:2012nq} 
  E.~Aprile {\it et al.}  [XENON100 Collaboration],
  Phys.\ Rev.\ Lett.\  {\bf 109}, 181301 (2012)
  [arXiv:1207.5988 [astro-ph.CO]].

\bibitem{DrlicaWagner:2012ry} 
  A.~Drlica-Wagner [Fermi LAT Collaboration],
  arXiv:1210.5558 [astro-ph.HE].

  \bibitem{ATLAS:2013pma} 
  [ATLAS Collaboration],
  ATLAS-CONF-2013-011.
  
\bibitem{Ellis:1975ap} 
  J.~R.~Ellis, M.~K.~Gaillard and D.~V.~Nanopoulos,
  Nucl.\ Phys.\ B {\bf 106}, 292 (1976).


\bibitem{Shifman:1979eb} 
  M.~A.~Shifman, A.~I.~Vainshtein, M.~B.~Voloshin and V.~I.~Zakharov,
  Sov.\ J.\ Nucl.\ Phys.\  {\bf 30}, 711 (1979)
  [Yad.\ Fiz.\  {\bf 30}, 1368 (1979)].
  
  \bibitem{Jora:2013opa} 
  R.~Jora, S.~Nasri and J.~Schechter,
  arXiv:1302.6344 [hep-ph].

\bibitem{Peskin:1990zt} 
  M.~E.~Peskin and T.~Takeuchi,
  Phys.\ Rev.\ Lett.\  {\bf 65}, 964 (1990).

\bibitem{Andreas:2008xy} 
  S.~Andreas, T.~Hambye and M.~H.~G.~Tytgat,
  JCAP {\bf 0810}, 034 (2008)
  [arXiv:0808.0255 [hep-ph]].

\bibitem{Poland:2008ev} 
  D.~Poland and J.~Thaler,
  JHEP {\bf 0811}, 083 (2008)
  [arXiv:0808.1290 [hep-ph]].
  
\bibitem{Bazzocchi:2012pp} 
  F.~Bazzocchi and M.~Fabbrichesi,
  Eur.\ Phys.\ J.\ C {\bf 73}, 2303 (2013)
  [arXiv:1207.0951 [hep-ph]].
  
\bibitem{Fitzpatrick:2010em} 
  A.~L.~Fitzpatrick, D.~Hooper and K.~M.~Zurek,
  Phys.\ Rev.\ D {\bf 81}, 115005 (2010)
  [arXiv:1003.0014 [hep-ph]].

\bibitem{Giedt:2009mr} 
  J.~Giedt, A.~W.~Thomas and R.~D.~Young,
  Phys.\ Rev.\ Lett.\  {\bf 103}, 201802 (2009)
  [arXiv:0907.4177 [hep-ph]].

\bibitem{Eidelman:2004wy} 
  S.~Eidelman {\it et al.}  [Particle Data Group Collaboration],
  Phys.\ Lett.\ B {\bf 592}, 1 (2004).

\bibitem{Craig:2013xia} 
  N.~Craig, C.~Englert and M.~McCullough,
  arXiv:1305.5251 [hep-ph].

  \bibitem{Belanger:2010pz}
  G.~Belanger, F.~Boudjema, A.~Pukhov and A.~Semenov,
  arXiv:1005.4133 [hep-ph].

\bibitem{Veltman:1980mj} 
  M.~J.~G.~Veltman,
  Acta Phys.\ Polon.\ B {\bf 12}, 437 (1981).

\bibitem{ATLAS:2013sla} 
  [ATLAS Collaboration],
  ATLAS-CONF-2013-034.
  
\bibitem{ATLAS:2012bmv} 
  [ATLAS Collaboration],
  ATLAS-CONF-2012-169.
  
\bibitem{CMS:yxa} 
  [CMS Collaboration],
  CMS-PAS-HIG-12-041.
  
  
\bibitem{ATLAS:2013ima} 
  [ATLAS Collaboration],
  ATLAS-CONF-2013-018.
  
\bibitem{ATLAS:2013qla} 
  [ATLAS Collaboration],
  ATLAS-CONF-2013-036.
      
\bibitem{ATLAS:2013rla} 
  [ATLAS Collaboration],
  ATLAS-CONF-2013-035.
  
\bibitem{atlas2013007}
[ATLAS Collaboration],
     ATLAS-CONF-2013-007.  
  

\bibitem{ATLAS:2013bma} 
  [ATLAS Collaboration],
  ATLAS-CONF-2013-025.
  
  
  \bibitem{Aad:2012bt} 
  G.~Aad {\it et al.}  [ATLAS Collaboration],
  Phys.\ Rev.\ D {\bf 86}, 012007 (2012)
  [arXiv:1202.3389 [hep-ex]].

\bibitem{atlas2013051}
[ATLAS Collaboration],
     ATLAS-CONF-2013-051.
     
  \bibitem{ATLAS:2012znl} 
  [ATLAS Collaboration],
  ATLAS-CONF-2012-168.

\bibitem{Alwall:2011uj} 
  J.~Alwall, M.~Herquet, F.~Maltoni, O.~Mattelaer and T.~Stelzer,
  JHEP {\bf 1106}, 128 (2011)
  [arXiv:1106.0522 [hep-ph]].

\bibitem{perelstein} 
  M.~Farina, M.~Perelstein and N.~R.~-L.~Lorier,
  arXiv:1305.6068 [hep-ph].

\bibitem{Sjostrand:2006za} 
  T.~Sjostrand, S.~Mrenna and P.~Z.~Skands,
  JHEP {\bf 0605}, 026 (2006)
  [hep-ph/0603175].
 
\bibitem{CMS:2011cra} 
  [CMS Collaboration],
  CMS-PAS-BTV-11-001.
  
\bibitem{Aad:2011mk} 
  G.~Aad {\it et al.}  [ATLAS Collaboration],
  Eur.\ Phys.\ J.\ C {\bf 72}, 1909 (2012)
  [arXiv:1110.3174 [hep-ex]].

\bibitem{talk} [CMS Collaboration]  
  https://twiki.cern.ch/twiki/pub/CMSPublic/PhysicsResultsMUO/SingleMuonEfficiencies\_2012.pdf


\end{thebibliography}
\renewcommand{\refname}{Bibliography}
\addcontentsline{toc}{section}{Bibliography}
\providecommand{\href}[2]{#2}\begingroup\raggedright

\end{document}